\documentclass[journal]{IEEEtran}

\usepackage[english]{babel}
\usepackage[T1]{fontenc}
\usepackage[utf8]{inputenc}
\usepackage{amsmath,amssymb,amsfonts,bm}
\usepackage{graphicx}
\usepackage{cite}
\usepackage{url}
\usepackage{booktabs}
\usepackage{algorithm,algpseudocode}
\usepackage{siunitx}
\usepackage[acronym,nonumberlist]{glossaries}
\usepackage{xcolor}
\usepackage{tikz}
\usetikzlibrary{matrix,calc,arrows.meta,positioning,angles,quotes,decorations.pathmorphing,shapes.geometric}
\usepackage{empheq}
\usepackage{subcaption}
\usepackage{comment}
\usepackage[hidelinks,bookmarks=false]{hyperref} 

\makeglossaries

\newacronym{3gpp}{3GPP}{3rd generation partnership project}
\newacronym{5g}{5G}{the 5th generation of mobile networks}
\newacronym{6g}{6G}{sixth-generation}
\newacronym{ack}{ACK}{acknowledgement}
\newacronym{afdm}{AFDM}{affine frequency division multiplexing}
\newacronym{aoi}{AoI}{age-of-information}
\newacronym{ap}{AP}{access point}
\newacronym{awgn}{AWGN}{additive white Gaussian noise}
\newacronym{b5g}{B5G}{beyond 5G}
\newacronym{ber}{BER}{bit-error rate}
\newacronym{bp}{BP}{belief propagation}
\newacronym{bs}{BS}{base station}
\newacronym{cscg}{CSCG}{circularly symmetric complex gaussian}
\newacronym{csi}{CSI}{channel state information}
\newacronym{dd}{DD}{delay–Doppler}
\newacronym{dir}{DIR}{data information rate}
\newacronym{dof}{DoF}{degrees-of-freedom}
\newacronym{drl}{DRL}{deep rl}
\newacronym{dtmc}{DTMC}{discrete-time markov chain}
\newacronym{embb}{eMBB}{enhanced mobile broadband}
\newacronym{fdma}{FDMA}{frequency division multiple access}
\newacronym{fec}{FEC}{forward error correction}
\newacronym{fifo}{FIFO}{first-in first out}
\newacronym{fim}{FIM}{fisher information matrix}
\newacronym{ga}{GA}{genetic algorithm}
\newacronym{gap}{GAP}{generalized assignment problem}
\newacronym{geo}{GEO}{geosynchronous earth orbit}
\newacronym{gsl}{GSL}{ground-to-satellite link}
\newacronym{icsi}{ICSI}{imperfect channel state information}
\newacronym{irsa}{IRSA}{irregular repetition slotted aloha}
\newacronym{isa}{ISA}{isolated spectral allocation}
\newacronym{isac}{ISAC}{integrated sensing and communication}
\newacronym{isic}{ISIC}{imperfect successive interference cancellation}
\newacronym{isl}{ISL}{inter-satellite link}
\newacronym{iot}{IoT}{internet of things}
\newacronym{leo}{LEO}{low earth orbit}
\newacronym{lmmse}{LMMSE}{linear minimum mean square error}
\newacronym{los}{LOS}{line-of-sight}
\newacronym{mab}{MAB}{multi-armed bandit}
\newacronym{mdp}{MDP}{markov decision process}
\newacronym{mgap}{MGAP}{multi-level generalized assignment problem}
\newacronym{mimo}{MIMO}{multiple-input multiple-output}
\newacronym{miso}{MISO}{multiple-input single-output}
\newacronym{mmimo}{mMIMO}{massive mimo}
\newacronym{mmse}{MMSE}{minimum mean square error}
\newacronym{mmtc}{mMTC}{massive machine-type communications}
\newacronym{mp}{MP}{message passing}
\newacronym{mu}{MU}{multi-user}
\newacronym{nack}{NACK}{negative ack}
\newacronym{noma}{NOMA}{non-orthogonal multiple access}
\newacronym{nr}{NR}{new radio}
\newacronym{ntn}{NTN}{non-terrestrial network}
\newacronym{ofdm}{OFDM}{orthogonal frequency division multiplexing}
\newacronym{ofdma}{OFDMA}{orthogonal frequency-division multiple access}
\newacronym{oma}{OMA}{orthogonal multiple access}
\newacronym{otfs}{OTFS}{orthogonal time frequency space}
\newacronym{pdf}{PDF}{probability density function}
\newacronym{pwf}{PWF}{phase warping function}
\newacronym{qos}{QoS}{quality of service}
\newacronym{ran}{RAN}{radio access network}
\newacronym{reir}{REIR}{radar estimation information rate}
\newacronym{rl}{RL}{reinforcement learning}
\newacronym{rms}{RMS}{root mean square}
\newacronym{rs}{RS}{rate-splitting}
\newacronym{rsma}{RSMA}{rate-splitting multiple access}
\newacronym{sagin}{SAGIN}{space–air–ground integrated network}
\newacronym{sc}{SC}{superposition coding}
\newacronym{sdma}{SDMA}{space-division multiple access}
\newacronym{sic}{SIC}{successive interference cancellation}
\newacronym{sinr}{SINR}{signal-to-interference-plus-noise ratio}
\newacronym{siso}{SISO}{single-input single-output}
\newacronym{snr}{SNR}{signal-to-noise ratio}
\newacronym{sp}{SP}{signal processing}
\newacronym{tacae}{TCAE}{time-averaged cost of actuation error}
\newacronym{tare}{TRE}{time-averaged reconstruction error}
\newacronym{tgp}{TGP}{tunable Gaussian pulse}
\newacronym{tdma}{TDMA}{time division multiple access}
\newacronym{tf}{TF}{time–frequency}
\newacronym{uav}{UAV}{unmanned aerial vehicle}
\newacronym{uc}{UC}{update-delivery cost}
\newacronym{ue}{UE}{user equipment}
\newacronym{urllc}{URLLC}{ultra-reliable and low-latency communications}
\newacronym{v2x}{V$2$X}{vehicle-to-everything}

\newacronym{af}{AF}{ambiguity function}
\newacronym{ftn}{FTN}{faster-than-Nyquist}
\newacronym{csit}{CSIT}{channel state information at the transmitter}
\newacronym{csir}{CSIR}{channel state information at the receiver}
\newacronym{acf}{ACF}{auto-correlation function}
\newacronym{crlb}{CRLB}{Cram\'er-Rao lower bound}
\newacronym{rrc}{RRC}{Root Raised Cosine}
\newacronym{ul}{UL}{uplink}
\newacronym{semi-isac}{Semi-ISaC}{Semi-Integrated Sensing and Communication}
\newacronym{cu}{CU}{communication user}
\newacronym{rt}{RT}{radar target}
\newacronym{rcs}{RCS}{radar cross-section}
\newacronym{dl}{DL}{downlink}
\newacronym{op}{OP}{Outage Probability}
\newacronym{wssus}{WSSUS}{wide-sense stationary uncorrelated scattering}
\newacronym{sgp}{SGP}{Standard Gaussian Pulse}
\newacronym{gbps}{Gbps}{Gigabits per second}
\newacronym{oobe}{OOBE}{Out-of-Band Emission}
\newacronym{papr}{PAPR}{peak-to-average power ratio}
\newacronym{ipr}{IPR}{Interference Power Ratio}
\newacronym{ici}{ICI}{Inter-Carrier Interference}
\newacronym{isi}{ISI}{Inter-Symbol Interference}
\newacronym{rr}{RR}{Radar Receiver}
\newacronym{dmrs}{DMRS}{Demodulation Reference Signals}
\newacronym{pusch}{PUSCH}{Physical Uplink Shared Channel}
\newacronym{cp}{CP}{Cyclic Prefix}
\newacronym{ifft}{IFFT}{Inverse Fast Fourier Transform}
\newacronym{fft}{FFT}{Fast Fourier Transform}
\newacronym{cfr}{CFR}{Channel Frequency Response }
\newacronym{dft}{DFT}{Discrete Fourier Transform}
\newacronym{dpi}{DPI}{Direct-Path Interference}
\newacronym{tx}{TX}{Transmitter}
\newacronym{1tfde}{1T-FDE}{One-Tap Frequency-Domain Equalization}
\newacronym{iota}{IOTA}{Isotropic Orthogonal Transform Algorithm}
\newacronym{io}{I/O}{Input/Output}
\newacronym{sfft}{SFFT}{Symplectic Finite Fourier Transform}
\newacronym{isfft}{ISFFT}{Inverse SFFT}
\newacronym{lct}{LCT}{Linear Canonical Transform}
\newacronym{oddm}{ODDM}{Orthogonal Delay-Doppler Division Multiplexing Modulation}  

\definecolor{cadmiumgreen}{rgb}{0.0, 0.42, 0.24}
\definecolor{ao(english)}{rgb}{0.0, 0.5, 0.0}

\newcommand{\bc}{\textcolor{black}} 
\newcommand{\am}{\textcolor{black}} 
\newcommand{\ilm}{\textcolor{black}} 
\newcommand{\ta}{\textcolor{black}} 
\newcommand{\pp}{\textcolor{black}} 

\newcommand{\erf}{\mathrm{erf}}

\def\BibTeX{{\rm B\kern-.05em{\sc i\kern-.025em b}\kern-.08em
    T\kern-.1667em\lower.7ex\hbox{E}\kern-.125emX}}

\begin{document}

\title{Tunable Gaussian Pulse for Delay-Doppler ISAC}

\author{\bc{Bruno~Felipe~Costa},~\IEEEmembership{Student Member,~IEEE,}
        \am{Anup~Mishra},~\IEEEmembership{Member,~IEEE,}
        \ilm{Israel~Leyva-Mayorga},~\IEEEmembership{Member,~IEEE,}
        \ta{Taufik~Abrão},~\IEEEmembership{Senior Member,~IEEE,}
        and~\pp{Petar~Popovski},~\IEEEmembership{Fellow,~IEEE}\vspace{-0.5cm}%
\thanks{Manuscript received Month Day, 2025; revised Month Day, 2025.}%
\thanks{B. F. Costa and T. Abrão are with the Department of Electrical Engineering, State University of Londrina (UEL), Londrina, PR, 86057-970, Brazil (e-mail: bruno.felipe.costa@uel.br; taufik@uel.br).}%
\thanks{A. Mishra, I. Leyva-Mayorga, and P. Popovski are with the Department of Electronic Systems, Aalborg University, 9220 Aalborg, Denmark (e-mail: anmi@es.aau.dk; ilm@es.aau.dk; petarp@es.aau.dk).}%
}


\maketitle
\glsresetall

\begin{abstract}
\Gls{isac} for next-generation networks targets robust operation under high mobility and high Dopplers spread, leading to severe \gls{ici} in systems based on \gls{ofdm} waveforms. \Gls{dd}-domain \gls{isac} offers a more robust foundation under high mobility, but it requires a suitable \gls{dd}-domain pulse shaping filter. The prevailing \gls{dd} pulse designs are either communication-centric or static, which limits adaptation to non-stationary channels and diverse application demands. To address this limitation, this paper introduces the \gls{tgp}, a \gls{dd}-native, analytically tunable pulse shape parameterized by its aspect-ratio ($\boldsymbol{\gamma}$), chirp-rate ($\boldsymbol{\alpha_c}$), and phase-coupling ($\boldsymbol{\beta_c}$). On the sensing side, we derive closed-form \glspl{crlb} that map $(\boldsymbol{\gamma,\alpha_c,\beta_c)}$ to fundamental delay and Doppler precision. On the communications side, we show that $\boldsymbol{\alpha_c}$ and $\boldsymbol{\beta_c}$ reshape off-diagonal covariance, and thus \gls{isi}, without changing  received power, isolating capacity effects to interference structure rather than power loss. A comprehensive trade-off analysis demonstrates that the \gls{tgp} spans a flexible operational region from the high capacity of the Sinc pulse to the high precision of the \gls{rrc} pulse. Notably, \gls{tgp} attains near-\gls{rrc} sensing precision while retaining over $90\%$ of Sinc’s maximum capacity, achieving a balanced operating region that is not attainable by conventional static pulse designs.
\end{abstract}


\glsresetall 

\vspace{-1em}
\section{Introduction}
\label{sec:introduction}
\IEEEPARstart{T}{he} vision for \gls{6g} networks is a unified platform where sensing and communications are natively integrated \cite{Yuan2021,Shuangyang2022,Mishra_coexistence}. This vision includes centimeter-level localization working alongside \gls{gbps} data links. In fast time-varying channels, \gls{ofdm}-based \gls{isac} suffers loss of subcarrier orthogonality under high Doppler, leading to \gls{ici} that degrades detection and ranging. In these regimes, \gls{dd}-domain \gls{isac} has emerged as a robust foundation, which aligns pulse design with radar estimation and enhances sensing precision~\cite{Yuan2024,Wie_OTFS}. Modulations such as \gls{otfs}, realized through Zak processing or \gls{sfft}/\gls{isfft} chain, operate on  \gls{dd} lattice where doubly dispersive channels become sparse and near-invariant \cite{Gopalam2024}. This inherent robustness is theoretically grounded in the \gls{dd} `crystallization condition', an operating regime in which the channel is essentially non-fading, thereby improving \gls{csi} coherence and simplifying equalization~\cite{Mohammed2023,Wie_OTFS}. Together, these properties make \gls{dd}-\gls{isac} a natural basis for joint perception and high-rate connectivity in high mobility scenarios\cite{Yuan2024}.
\par Crucially, one of the key imperatives for realizing the performance potential of \gls{dd}-\gls{isac} is \emph{\gls{dd}-domain pulse shaping}~\cite{Yuan2024,Li2023}. This entails specifying the pulse’s envelope, i.e., its delay and Doppler spread, and its phase structure so that sensing attains sharp delay–Doppler distinction (as quantified by standard ambiguity metrics) while \gls{isi} is controlled for reliable communication~\cite{Li2023,Liu2025iceberg}. Consequently, the pulse jointly determines sensing precision and the conditioning of the effective \gls{dd}-domain channel that supports data detection~\cite{liu2025pulse}. Motivated by the above, this paper focuses on \gls{dd}-domain pulse shaping to jointly enhance sensing and communication performance in \gls{dd}-\gls{isac}.
\vspace{-1em}
\subsection{Related Work}
Over the years, the pulse shape has shifted from an implementation detail to a central design concern in \gls{dd}-domain systems~\cite{Mohammed2023,Jayachandran2024,Liu2025}. Unlike conventional time-domain filtering, \gls{dd} pulse shaping directly sculpts the fundamental \gls{isac} trade-off by controlling two critical and often conflicting aspects \cite{Yuan2024,Li2023}. On the sensing side, the pulse’s \gls{af} structure dictates target resolution and ranging accuracy~\cite{Shen2022,Kafedziski2025}, \bc{and recent advances~\cite{Liu2025iceberg} demonstrate that the deterministic pulse geometry emerges as the fundamental sensing limit once random data interference is suppressed by coherent integration, thereby binding the ultimate range performance to the pulse's autocorrelation properties.} On the communication side, reliability is governed by \gls{isi} and \gls{oobe}, which control \gls{ber} saturation at high \gls{snr}~\cite{Jayachandran2024,Shen2022}.
\par Subsequently, the research community has explored various pulse shapes. Recent work, including static filters like the Gaussian-Sinc~\cite{Das2025a} and systematic design frameworks based on Hermite functions~\cite{Jesbin2025} or the \gls{iota}~\cite{Mehrotra2025}, has focused on   \emph{communication-centric} \gls{isac} trade-off. In these works, sensing is primarily defined as the task of \gls{io} relation estimation (i.e., channel estimation) for the benefit of the communication task. The optimization goal is therefore to balance communication \gls{isi} (favouring full, near, or bi-orthogonality) against \gls{io} estimation accuracy, which requires low \gls{af} sidelobes to prevent \gls{dd} aliasing~\cite{Das2025a, Jesbin2025, Mehrotra2025}. Within this sidelobe-suppression lens, the standard Gaussian pulse is correctly identified as a strong candidate for sensing due to its inherently sidelobe-free \gls{af}; in finite or periodic realizations it still exhibits comparatively low sidelobes.~\cite{Jayachandran2024, Kafedziski2025}. While these advances represent significant progress in \gls{dd}-domain pulse design, a critical examination reveals \pp{three} fundamental limitations when considering a broader \emph{radar-centric} \gls{isac} design space.

\par The \emph{first} limitation is methodological. A sensing metric based solely on sidelobe suppression is insufficient for radar-centric tasks, where precision is governed by the \gls{crlb} and depends on the \gls{af}'s local peak curvature~\cite{Dogandzic2001, Iain2006}. \pp{This reveals a trade-off: the very smoothness that endows the Gaussian pulse with desirable sidelobe suppression simultaneously produces a rounded \gls{af} peak with insufficient curvature, rendering it intrinsically suboptimal for high-precision delay-Doppler estimation~\cite{Iain2006}.} The \emph{second} limitation is practical, \pp{as} the reliance on static pulses~\cite{Das2025a} prevents adaptation to non-stationary channel conditions that vary dramatically across scenarios, from highway vehicular networks (dominated by Doppler spread) to dense urban environments (dominated by delay spread)~\cite{Pfadler2020}. Finally, the \emph{third} limitation concerns communication. While resource allocation strategies have been extensively studied~\cite{Xinyang2023, Mura2025}, the fundamental question of how pulse \emph{phase structure} at the transmitter shapes the statistical interference pattern and, consequently, the achievable capacity, remains significantly under-explored. Thus, a clear and significant gap remains: there is no \gls{dd}-native, analytically \emph{tunable} pulse shape designed to navigate the true \gls{isac} trade-off: communication capacity~\cite{Jiao2025} versus radar estimation precision~\cite{Gaudio2020a, Chen2021}.
\vspace{-1em}
\subsection{Contributions}
To overcome the limitations and fill the research gap discussed above, this paper introduces the \gls{tgp}, a novel \gls{dd}-domain pulse shape designed from first principles for \gls{isac} co-design. The \gls{tgp} is parametrised by three intuitive and physically meaningful knobs that provide explicit analytical control over envelope geometry and phase structure. Our main contributions are: 
\begin{enumerate}
    \item We propose the \gls{tgp}, a novel, analytically tractable, and continuously tunable \gls{dd}-domain pulse. Its parameters, the aspect-ratio ($\gamma$), chirp-rate ($\alpha_c$), and phase-coupling ($\beta_c$), offer explicit control over the pulse's geometry and phase structure, enabling fine-grained optimization of the \gls{isac} trade-off beyond the static designs~\cite{Das2025a, Jesbin2025, Mehrotra2025}. \pp{This \gls{dd}-domain pulse shaping design enables dynamic adaptation to joint sensing and communication requirements through explicit analytical tuning\footnote{\bc{In contrast to the scalar roll-off parameter of the \gls{rrc} used for spectral compliance, \gls{tgp} parameters shape the delay-Doppler coupling to jointly optimize the sensing and communication functionalities.}}.}

    \item We derive the closed-form expressions for the \gls{fim} and the resulting \gls{crlb} for delay–Doppler estimation. These expressions provide an explicit mapping from tuning parameters $(\gamma,\alpha_c,\beta_c)$ to the fundamental limits of sensing precision~\cite{Dogandzic2001, Iain2006}, addressing the methodological gap left by sidelobe-focused sensing metrics~\cite{Jayachandran2024, Kafedziski2025}.

    \item We provide a comprehensive communication performance analysis by deriving the effective channel's second-order statistics under a \gls{wssus} channel model. Crucially, we prove analytically that the \gls{tgp}'s phase parameters ($\alpha_c, \beta_c$) reshape the channel's interference structure via off-diagonal covariance elements without altering received power. This result establishes a crucial link between transmitter-side pulse phase design and achievable ergodic capacity, demonstrating a novel decoupling mechanism  within the \gls{dd}-domain. 

    \item We characterize the achievable \gls{isac} trade-off landscape by numerically evaluating the radar \gls{crlb} and communication capacity across an extensive parameter sweep of the \gls{tgp}'s tuning knobs ($\gamma,\alpha_c,\beta_c$). We benchmark \gls{tgp} against well-studied static pulses, i.e., Sinc, \gls{rrc}, and \gls{sgp}. To this end, we move beyond the \gls{wssus} channel model and use the 3GPP Vehicular-A (Veh-A) six-path channel for the communication user, a standard high-mobility benchmark in pulse-shaping studies~\cite{Mehrotra2025,Jesbin2025,Das2025a,ITUM1225}. Through numerical results, we demonstrate that the \gls{tgp} \bc{mathematically generalizes the \gls{sgp} and} traces a tunable Pareto frontier that envelops the fixed operating points of \bc{the other} baselines, bridging the gap between sensing-centric and communication-centric designs \bc{in a domain previously limited to static pulse shapes.}
    
\end{enumerate}
\vspace{-1em}
\subsection{Organisation}

The remainder of this paper is organized as follows.
Section~\ref{sec:system_model} details the proposed system model, introducing the analytically tractable \gls{tgp} and formulating the effective \gls{dd} channel for both sensing and communication. 
Section~\ref{sec:performance_analysis} provides a detailed communication performance analysis, where we derive the channel's second-order statistics to establish an analytical link between the \gls{tgp}'s phase parameters and the ergodic capacity. 
In Section~\ref{sec:sensing_performance}, we shift the focus to fundamental sensing performance, deriving the closed-form \gls{fim} and the corresponding \gls{crlb} to quantify estimation precision. 
Section~\ref{sec:numerical_results} presents extensive \gls{isac} numerical results, benchmarking the \gls{tgp} against state-of-the-art pulses and characterizing the \gls{isac} trade-off landscape it enables. 
Finally, Section~\ref{sec:Conclusion} concludes the paper.
\vspace{-1em}
\subsection{{Notation}}
\label{subsec:notation}

Scalars are denoted by lowercase italic (\(x\)); vectors by bold lowercase (\(\mathbf{x}\)); vectors by bold italic (e.g., \(\boldsymbol{\theta}\)); and matrices by bold uppercase (\(\mathbf{X}\)). The sets \(\mathbb{R}\), \(\mathbb{C}\), and \(\mathbb{Z}\) denote real numbers, complex numbers, and integers, respectively, \(\mathcal{CN}(\boldsymbol{\mu},\mathbf{\Sigma})\) denotes a \gls{cscg} distribution with mean \(\boldsymbol{\mu}\) and covariance \(\mathbf{\Sigma}\), and $\mathcal{U}[a, b)$ denotes the continuous uniform distribution over the interval $[a, b)$. We use \((\cdot)^*\) for complex conjugate, \((\cdot)^{\mathrm{H}}\) for Hermitian transpose, \(\mathbb{E}[\cdot]\) for expectation, \(\det(\cdot)\) for determinant, and \(\Re\{\cdot\}\), \(\Im\{\cdot\}\) for real and imaginary parts. The symbols \(\delta(\cdot)\) and \(\erf(\cdot)\) denote the Dirac delta and the error function, symbol \( *_{\sigma} \) denotes twisted convolution, and \(\mathrm{mod}\) denotes modular arithmetic. Domain-specific objects used throughout are the ambiguity function \(\mathrm{AF}(\cdot)\), the Fisher information matrix \(\mathbf{I}(\cdot)\), and the Cramér–Rao lower bound \(\mathrm{CRLB}(\cdot)\).

\vspace{-0.5em}
\section{System Model}
\label{sec:system_model}
\bc{We consider a communication and sensing system operating over a doubly-selective channel, characterized by both time and frequency dispersion (i.e., high Doppler and delay spreads). We model the underlying physical channel as a \gls{wssus} process. This section details the operational protocol designed for this environment, the \gls{dd}-domain pulse, and the resulting effective channel models.}

We consider a frame of duration $T$ over bandwidth $B$ on a \gls{dd} grid of size $M\times N$ with resolutions $\Delta\tau = T/M$ and $\Delta\nu = B/N$. Moreover, the transmit \gls{dd} pulse is normalized to unit energy; total symbol energy $E_s$ appears explicitly in \gls{snr}/\gls{fim} expressions as a pre-factor. Furthermore, for sensing, the complex channel gain $h_T$ is treated as known (or as a nuisance separated from the kinematic parameters) to isolate the impact of pulse geometry on estimation precision. For communication, analysis assumes perfect \gls{csir} of the \emph{effective} \gls{dd}-domain channel $\mathbf{H}$ (a standard benchmark assumption for capacity-style evaluations), while no \gls{csit} is used and power is allocated uniformly over \gls{dd} symbols.

\bc{\textit{Operational Protocol and Practical Considerations:} We adopt a protocol based on the separation of timescales, consistent with the \gls{wssus} pulse design framework~\cite{Jung2007}. Specifically, the \gls{tgp} parameters $(\gamma, \alpha_c, \beta_c)$ are tuned based on the channel's slowly-varying second-order statistics, while data transmission occurs over fast-fading realizations. This framework entails three key implications:
\begin{enumerate}
    \item \textit{Ergodic Regime:} Since codewords span multiple realizations of the effective channel $\mathbf{H}$ while the pulse remains fixed, the ergodic capacity serves as the performance metric. This justifies our focus on ergodic capacity rather than outage probability.
    \item \textit{Computational Complexity:} Pulse optimization is performed infrequently (e.g., via lookup tables based on statistical sensing). This implies that during runtime, the \gls{tgp} operates as a static filter with computational complexity identical to that of standard static pulses, thereby avoiding any real-time optimization overhead.
    \item \textit{Robustness:} While perfect \gls{csir} is assumed for deriving theoretical bounds, the reliance on stable, slowly-varying second-order statistics for tuning ensures that the \gls{tgp} design is resilient to the instantaneous estimation errors that can affect systems relying on perfect, real-time \gls{csi}.
\end{enumerate}}
\vspace{-1em}
\subsection{\Gls{dd} Domain Pulse}
%
The cornerstone of our work is a novel, adaptable \gls{dd}-domain pulse whose continuous representation is the product of a real Gaussian envelope and a complex phase term, given by:
\begin{equation}
    x_{\mathrm{dd}}(\tau,\nu)
    = \exp\!\left(-\pi\left(\frac{2\tau^2}{\gamma T^2} + \frac{2\gamma\nu^2}{B^2}\right)\right)\,\Psi(\tau,\nu).
    \label{eq:tgp_waveform_explicit}
\end{equation}
Here, $(\tau, \nu)$ denote the continuous delay (in s) and Doppler (in Hz) variables, respectively, over $T$ and $B$. The complex term $\Psi(\tau,\nu)$, which we refer to as the \gls{pwf}, is written as:
\begin{equation}
    \Psi(\tau,\nu) = \exp\!\left(j\pi\left(\frac{\alpha_c}{T^2}\tau^2 + \beta_c\,\tau\nu\right)\right).
    \label{eq:Phase_Warping_Function}
\end{equation}
Together, the dimensionless parameter set $(\gamma, \alpha_c, \beta_c)$ provides fine-grained control over the pulse structure. The geometry of the Gaussian envelope is dictated by the \emph{aspect-ratio parameter} $\gamma$, directly trading delay resolution for Doppler resolution. The \emph{chirp-rate parameter} $\alpha_c$ introduces a quadratic phase (a chirp) that acts as a switch to enable and set the strength of the delay-Doppler coupling. Finally, the \emph{phase-coupling parameter} $\beta_c$ applies a \emph{bilinear shear}, a term rooted in the theory of \gls{lct} where bilinear phase terms such as $\tau\nu$ induce geometric shearing in phase-space representations~\cite{Ozaktas2001}. This operation globally adjusts both the axial curvatures of the \gls{af} peak and the coupling strength. The analysis of the \gls{fim} in Section~\ref{sec:sensing_performance} will rigorously validate this division of roles.
\vspace{0.1cm}
\paragraph{Aspect-ratio parameter ($\gamma$)}
We parameterize the Gaussian envelope such that the \gls{rms} spreads satisfy
\begin{equation}
\sigma_{\tau}=c_{\tau}\sqrt{\gamma}\,T,\qquad
\sigma_{\nu}=c_{\nu}\frac{B}{\sqrt{\gamma}},
\end{equation}
so the delay–Doppler product \(\sigma_{\tau}\sigma_{\nu}=(c_{\tau}c_{\nu})\,BT\) is independent of $\gamma$. The constants $(c_{\tau},c_{\nu})$ depend only on the chosen \gls{rms}-moment convention; selecting $c_{\tau}c_{\nu}=1/(4\pi)$ aligns the family with the Heisenberg benchmark (up to convention), see~\cite{Richards2010}. Consequently, $\gamma$ controls the envelope aspect ratio and thus trades delay and Doppler resolution while preserving the total “information footprint” (for fixed $B$ and $T$): increasing $\gamma$ broadens the delay spread and tightens the Doppler spread (improving Doppler resolution), whereas smaller $\gamma$ does the opposite (favouring delay resolution).
\vspace{0.1cm}
\paragraph{Chirp–rate parameter ($\alpha_c$)}
The parameter $\alpha_c$ scales the quadratic phase term in the \gls{pwf}, thereby introducing a controlled frequency sweep (chirp) along the delay axis and breaking the pulse symmetry. At $\alpha_c=0$ the pulse is unchirped and the local \gls{af} peak is axis-aligned; for $\alpha_c\neq 0$ the induced quadratic phase injects delay–Doppler coupling, which appears as a nonzero mixed curvature in the \gls{af} and as a nonzero off-diagonal entry $I_{\tau\nu}$ in the \gls{fim} ($\mathbf{I}$). The sign and magnitude of $\alpha_c$ set the tilt direction and coupling strength, respectively, while the envelope aspect ratio remains governed by $\gamma$.
\vspace{0.1cm}
\paragraph{Phase–coupling parameter ($\beta_c$)}
The parameter $\beta_c$ scales the bilinear term in the \gls{pwf}, which imposes a shear in the $(\tau,\nu)$ phase space~\cite{Ozaktas2001} and, consequently, a global re-weighting of the \gls{af} curvatures. When $\beta_c=0$ the phase is purely quadratic (set by $\alpha_c$), and the delay and Doppler axes remain uncoupled in phase; turning on $\beta_c\neq 0$ modifies the axial curvatures and scales the mixed curvature induced by $\alpha_c$. In \gls{fim} terms, $\beta_c$ reshapes the diagonal entries $(I_{\tau\tau},I_{\nu\nu})$ and {modulates} the off-diagonal coupling $I_{\tau\nu}$, whereas $\alpha_c$ acts as the primary {switch} for coupling (indeed, $I_{\tau\nu}=0$ for $\alpha_c=0$ regardless of $\beta_c$). Section~\ref{sec:sensing_performance} provides the closed-form expressions and operating ranges ensuring positive definiteness of the \gls{fim}.
\vspace{0.1cm}
\paragraph{Positioning and scope}
\bc{The proposed \gls{tgp} is a tunable \gls{dd}-domain pulse shaping filter whose parameters $(\gamma,\alpha_c,\beta_c)$ control the envelope geometry and phase, and it reduces to the standard separable Gaussian pulse when $\alpha_c=\beta_c=0$. In this paper, we study \gls{tgp} within a pulse-shaped \gls{otfs}/Zak-\gls{otfs}-style architecture, and benchmark it against common static choices (Gaussian, Sinc, and \gls{rrc}). More broadly, \gls{tgp} is compatible with \gls{dd}-domain multicarrier frameworks that explicitly admit configurable transmit/receive pulses (or equivalent windowing) in their input--output model. Integrating \gls{tgp} into modulation schemes whose performance relies on a specifically constructed orthogonal pulse or chirp-eigenfunction basis, e.g., \gls{oddm}~\cite{Lin2022} or \gls{afdm}~\cite{Bemani2021}, would require a separate analysis to preserve their orthogonality/I--O structure, and is therefore left for future work.}
%
\vspace{-1em}
\subsection{Single-User Communication Model}
\label{subsec:comm_model}
%
We assess communication performance in a canonical single-user, point-to-point setting. A frame of $MN$ data symbols $x[k,l]$, drawn from a zero-mean, unit-variance constellation, is placed on the \gls{dd} grid and stacked into $\mathbf{x}\in\mathbb{C}^{MN}$. Each information symbol $x[k,l]$ modulates a Dirac impulse at the corresponding grid point $[k,l]$ in the \gls{dd} plane, forming an impulse train. This impulse train is filtered via \textit{twisted convolution} with the transmit pulse $x_{\mathrm{dd}}(\tau,\nu)$ defined in~\eqref{eq:tgp_waveform_explicit}. The resulting continuous \gls{dd} signal is synthesized into the time-domain waveform via the inverse Zak transform~\cite{Mohammed2023}.

The signal propagates through a doubly dispersive channel and is processed at the receiver via the Zak transform, yielding the discrete \gls{dd}-domain observation
\begin{equation}
    \mathbf{y} = \mathbf{H}\,\mathbf{x} + \mathbf{n},
    \label{eq:comm_rx_signal}
\end{equation}
where the data vector $\mathbf{x}\sim\mathcal{CN}(\mathbf{0},\mathbf{Q})$ with $\mathbf{Q}=\frac{E_s}{MN}\mathbf{I}$, and $\mathbf{n}\sim\mathcal{CN}(\mathbf{0},\sigma_n^2\mathbf{I})$ represents the \gls{dd}-domain \gls{awgn}. The structure of $\mathbf{H}$, in particular, its sparsity pattern, off-diagonal decay, and conditioning, is determined by the pulse shape $x_{\mathrm{dd}}(\tau,\nu)$ and the specific realization of the random physical channel. Section~\ref{subsec:effective_channel} provides the explicit construction of $\mathbf{H}$ and shows how $(\gamma,\alpha_c,\beta_c)$ control \gls{isi} and the resulting detection difficulty. As aforemention, we assume perfect \gls{csir} with respect to the effective channel matrix $\mathbf{H}$~\cite{Mishra2022a}.
\vspace{-0.5em}
\subsection{\gls{dd}-Pulse–Shaped Effective Channel}
\label{subsec:effective_channel}

The effective channel matrix $\mathbf{H}\in\mathbb{C}^{MN\times MN}$ is the discrete \gls{dd}-domain representation of the end-to-end operation, capturing how the \gls{dd} pulse $x_{\mathrm{dd}}(\tau,\nu)$ interacts with the physical multipath channel.
\vspace{0.1cm}
\paragraph{Physical channel}
We extend the monostatic sensing model from Sec.~\ref{subsec:sensing_model} to a general multipath scenario with $P$ propagation paths, each characterized by complex gains $h_i$, delays $\tau_i$, and Dopplers $\nu_i$\cite{Raviteja2019} 
\begin{equation}
    h(\tau,\nu)=\sum_{i=1}^{P} h_i \,\delta(\tau-\tau_i)\,\delta(\nu-\nu_i).
    \label{eq:physical_channel_continuous}
\end{equation}
Following the twisted-convolution operation described in~\eqref{eq:mean_echo_signal}, each path $i$ contributes a continuous \gls{dd}-domain response of the form
\begin{equation}
    y_i(\tau,\nu)=h_i\,x_{\mathrm{dd}}(\tau-\tau_i,\nu-\nu_i)\,e^{j2\pi(\nu\tau_i-\nu_i\tau)},
    \label{eq:single_path_response}
\end{equation}
which is a shifted, phase-rotated replica of the \gls{dd} pulse, analogous to the single-target echo in~\eqref{eq:mean_echo_signal} with $h_T\to h_i$, $\tau_T\to\tau_i$, and $\nu_T\to\nu_i$.

\vspace{0.1cm}

\paragraph{Discretization and Indices}
We define the discrete \gls{dd} grid indices. A transmitted symbol at index $q \equiv (m,n)$ is mapped to a received grid point at index $p \equiv (k,l)$, where $m, k \in \{0, \dots, M-1\}$ and $n, l \in \{0, \dots, N-1\}$. The analysis is based on a circular (mod-$M,N$) model of the finite Zak grid \cite{Gopalam2024}, which uses the resolutions $\Delta\tau$ and $\Delta\nu$ established at the beginning of this section.

For a given channel path $i$ with delay $\tau_i$ and Doppler $\nu_i$, we use an on-grid approximation where its corresponding discrete indices are $(k_i, l_i)$. The effective displacement between the transmitted and received symbol indices due to this path is then given by:
\begin{equation}
\begin{split}
k_{\text{diff}}&=(k-m-k_i)\bmod M,\\
l_{\text{diff}}&=(l-n-l_i)\bmod N.
\end{split}
\end{equation}
These offsets parameterise the path-induced shift across the grid; they enter the \gls{dd}-pulse spreading term and yield the element wise form of $\mathbf{H}_i$ in \eqref{eq:matrix_element_final}.

\vspace{0.1cm}
\paragraph{Superposition over paths}
The full effective channel is the superposition of individual paths, given by:
\begin{equation}
    \mathbf{H}(\gamma,\alpha_c,\beta_c)=\sum_{i=1}^{P}\mathbf{H}_i(\gamma,\alpha_c,\beta_c),
    \label{eq:full_channel_matrix_sum}
\end{equation}
%

\begin{figure*}[!htbp]
\begin{equation}
\footnotesize
\boxed{
[\mathbf{H}_i]_{p,q}
=
\underbrace{h_i}_{\text{(i) Path gain}}
\ \underbrace{\exp\!\Bigg(
-\pi\!\left[\frac{2(k_{\text{diff}}\Delta\tau)^2}{\gamma T^2}+\frac{2\gamma(l_{\text{diff}}\Delta\nu)^2}{B^2}\right]
+j\pi\!\left[\frac{\alpha_c}{T^2}(k_{\text{diff}}\Delta\tau)^2+\beta_c(k_{\text{diff}}\Delta\tau)(l_{\text{diff}}\Delta\nu)\right]
\Bigg)}_{\text{(ii) \gls{dd}-pulse spreading value at displacement }(k_{\text{diff}},l_{\text{diff}})}
\ \underbrace{\exp\!\left\{\,j2\pi\!\left(\frac{n\,l_i}{N}-\frac{m\,k_i}{M}\right)\right\}}_{\text{(iii) Discrete phase twist}}
}
\label{eq:matrix_element_final}
\end{equation}
\hrulefill
\end{figure*}


where $[\mathbf{H}_i]_{p,q}$ maps the input symbol at $q=(m,n)$ to the output sample at $p=(k,l)$ via the three multiplicative terms in \eqref{eq:matrix_element_final}: (i) the physical path gain $h_i$; (ii) the \gls{dd}-pulse spreading value at the displacement between $(k,l)$ and the path-shifted $(m,n)$; and (iii) the grid-induced phase twist that arises from the finite Zak grid representation.
\vspace{0.1cm}
\paragraph{Remark on Off-Grid Paths}
The preceding analysis assumes channel paths align perfectly with the \gls{dd} grid. In the practical case where $(\tau_i,\nu_i)$ are fractional (off-grid), the discrete path indices $(k_i,l_i)$ are replaced by their continuous counterparts. This means the \gls{dd}-pulse spreading term in \eqref{eq:matrix_element_final} samples the pulse's continuous \gls{af} at sub-bin displacements. While the resulting matrix $\mathbf{H}_i$ becomes dense, it retains the same structured decay, reinforcing the fundamental link between our pulse design and the resulting interference patterns.

\vspace{-0.5em}
\subsection{Monostatic Sensing Model}
\label{subsec:sensing_model}

We consider monostatic sensing of a single point target concurrent with data transmission. The target is characterized by its round-trip delay $\tau_T$, Doppler shift $\nu_T$, and channel gain $h_T$. The primary sensing task is to estimate the kinematic parameters $\bm{\theta}=[\tau_T,\nu_T]^T$. To derive fundamental performance limits, we assume $h_T$ is known, and focus on the impact of the pulse geometry on delay-Doppler estimation precision.

The \gls{bs} transmits a signal synthesized from the \gls{dd}-domain pulse $x_{\mathrm{dd}}(\tau,\nu)$ defined in~\eqref{eq:tgp_waveform_explicit}. The \gls{dd} impulse response of a single specular path is given as \cite{mohammed2024otfs}.
\begin{equation}
    h_{\text{target}}(\tau,\nu) = h_T\,\delta(\tau-\tau_T)\,\delta(\nu-\nu_T).
\label{eq:sensing_channel_impulse_response}
\end{equation}
The mean (noise-free) echo is obtained by the \textit{twisted convolution}\footnote{Twisted convolution between $a(\tau,\nu)$ and $b(\tau,\nu)$ is defined as $(a *_\sigma b)(\tau,\nu) = \iint a(\tau',\nu') b(\tau-\tau',\nu-\nu') e^{j2\pi\nu'\tau} d\tau' d\nu'$~\cite{mohammed2024otfs}. Note that it is non-commutative in nature.}  between $x_{\mathrm{dd}}$ and $h_{\text{target}}$; for a single path this yields a scaled, shifted, phase-rotated replica of the \gls{dd} pulse~\cite{mohammed2024otfs}
\begin{equation}
    \mu(\tau,\nu;\bm{\theta})
    = h_T\,x_{\mathrm{dd}}(\tau-\tau_T,\nu-\nu_T)\,e^{j2\pi(\nu\tau_T-\nu_T\tau)}.
    \label{eq:mean_echo_signal}
\end{equation}
We normalize the \gls{dd} pulse to unit energy,  and account for the total transmit energy $E_s$ in the \gls{snr} pre-factor used in Sec.~\ref{sec:sensing_performance}. This separation makes pulse comparisons fair  while keeping \gls{snr} bookkeeping explicit.
We model the disturbance on the \gls{dd} plane as zero-mean \gls{cscg} \gls{awgn},
\begin{equation}
    \mathbb{E}\!\left\{w(\tau,\nu)\,w^*(\tau',\nu')\right\}
    = N_0\,\delta(\tau-\tau')\,\delta(\nu-\nu'),
    \label{eq:dd_noise_psd}
\end{equation}
so, the observed echo is 
{\cite{Kay1993}}
\begin{equation}
    z(\tau,\nu) = \mu(\tau,\nu;\bm{\theta}) + w(\tau,\nu).
    \label{eq:received_echo}
\end{equation}
Upon sampling on an $M\times N$ \gls{dd} grid with spacings $(\Delta\tau,\Delta\nu)$, the discrete noise samples become i.i.d.\ with variance $\sigma_n^2 = {N_0}/(\Delta\tau\,\Delta\nu)$, ensuring consistency between the continuous model and the discrete receiver used in simulations.
\par The sensing task is to estimate $(\tau_T,\nu_T)$ from $z(\tau,\nu)$. The fundamental precision limits are governed by the local geometry of the \gls{dd} ambiguity induced by the transmitted \gls{dd} pulse $x_{\mathrm{dd}}(\tau,\nu)$ and are therefore directly controllable via the tunable parameters $(\gamma,\alpha_c,\beta_c)$; Section~\ref{sec:sensing_performance} formalizes this connection through the \gls{fim}.

\section{Communication Performance Analysis}
\label{sec:performance_analysis}

We evaluate the intrinsic quality of the effective \gls{dd}-domain channel $\mathbf{H}$ induced by the interaction between the \gls{dd} pulse (\gls{tgp}) and the physical channel. Because the \gls{tgp} creates structured, non-orthogonal interference in the \gls{dd} grid (akin to \gls{ftn}, type signaling), we adopt the \bc{ergodic capacity} with Gaussian inputs as the primary, receiver-agnostic performance metric, \bc{following the framework in} \cite{Zhang2025Capacity}. 
\bc{Consistent with the fast-fading assumption in Sec.~\ref{sec:system_model}, the ergodic capacity is defined as the expectation of the instantaneous mutual information over the ensemble of channel realizations.} Using \eqref{eq:comm_rx_signal} with perfect \gls{csir} and defining \(\mathrm{SNR}=E_s/\sigma_n^2\), and with equal power across the \(MN\) DD symbols, this \bc{Gaussian-input instantaneous mutual information} for a given realization $\mathbf{H}$ is 
\begin{equation}
    \bc{I}(\mathbf{H}) = \log_2 \det\left(\mathbf{I} + \frac{\text{SNR}}{MN} \mathbf{H} \mathbf{H}^H\right) \,[\text{bits/frame}].
    \label{eq:channel_capacity}
\end{equation}
\par We adopt uniform power allocation across all \gls{dd} symbols, which is optimal when no \gls{csit} is available~\cite{Palomar2003}. This ensures \bc{that $I(\mathbf{H})$} in \eqref{eq:channel_capacity} depends solely on the singular value spectrum of the effective channel matrix $\mathbf{H}$, which the \gls{tgp} parameters $(\gamma,\alpha_c,\beta_c)$ directly control: increasing axial localization (via $\gamma$) concentrates energy near the main diagonal, reducing \gls{isi}, while phase warping (via $\alpha_c,\beta_c$) reshapes the off-diagonal structure and eigenvalue spread.

\vspace{-1em}
\subsection{Analytical Framework via Channel Covariance}
\label{sub:analytical_framework}
To forge a direct analytical link between the \gls{tgp} parameters and the statistical communication performance, we focus on the \bc{ergodic capacity}, given by the expectation \(C_{\mathrm{erg}}=\mathbb{E}[\bc{I}(\mathbf{H})]\). Since the log-det($\cdot$) expectation is intractable, we invoke Jensen’s inequality to obtain a tractable upper bound used as our analytic proxy.
\begin{equation}
    C_{\mathrm{erg}} \le C_{\mathrm{jensen}} = \log_2 \det\left(\mathbf{I} + \frac{\text{SNR}}{MN} \mathbb{E}[\mathbf{H}\mathbf{H}^H]\right).
    \label{eq:jensen_bound}
\end{equation}
\bc{Numerical validation confirms the tightness of this bound for the considered scenarios, yielding a deviation of less than $1\%$ relative to the Monte Carlo simulated ergodic capacity.} This strategic shift transforms the problem from analyzing a random matrix instance to characterizing its second-order statistics. Our central task becomes the derivation of the average channel covariance matrix, $\mathbf{R_H} \triangleq \mathbb{E}[\mathbf{H}\mathbf{H}^H]$. 

We assume a \gls{wssus} channel model, where the $P$ propagation paths are statistically independent and uniformly distributed; this represents a canonical, analytically tractable model of a richly scattered channel, consistent with foundational pulse design literature~\cite{Jung2007}. Let $\mathbf{H}^{(i)}$ be the effective channel matrix corresponding to the $i$-th path. The entry of the average channel covariance matrix $\mathbf{R_H}$ mapping an input symbol at index $k$ to an output symbol at index $j$ 
\begin{equation}
    [\mathbf{R_H}]_{j, k} = P \sum_{\ell=0}^{MN-1} \mathbb{E}\left[H_{j,\ell}^{(1)} \left(H_{k,\ell}^{(1)}\right)^*\right],
    \label{eq:rh_sum_expanded}
\end{equation}
where the expectation is taken over the random parameters of a single path: its complex gain $h_1$ (with variance $\sigma_h^2$), delay $\tau_1 \in [0, \tau_{\max}]$, and Doppler $\nu_1 \in [-\nu_{\max}, \nu_{\max}]$. The core of our analysis is to find a closed-form solution for this expectation, whose structure is dictated by the complex phase of the \gls{tgp}-shaped channel coefficient.

Let the complex exponent of a single matrix element $H_{j,\ell}^{(1)}$ be denoted $\phi(\tau, \nu; j, \ell)$. This exponent comprises a real part (log-magnitude) and an imaginary part (phase):
\begin{align}
    \phi(\tau, \nu; j, \ell) &= \phi_{\text{real}}(\tau, \nu; j, \ell) + j\phi_{\text{imag}}(\tau, \nu; j, \ell), \label{eq:phi_components}
\end{align}
where, with $\Delta\tau = (k-m)\Delta\tau - \tau$ and $\Delta\nu = (l-n)\Delta\nu - \nu$, we have
\begin{align}
\phi_{\text{real}} &= -\pi\left( \frac{2\Delta\tau^2}{\gamma T^2} + \frac{2\gamma\Delta\nu^2}{B^2} \right), \label{eq:phi_real_def} \\
\phi_{\text{imag}} &= \pi\left( \frac{\alpha_c}{T^2}\Delta\tau^2 + \beta_c\Delta\tau\Delta\nu \right) + \frac{2\pi l \nu}{N\Delta\nu} - \frac{2\pi m \tau}{M \Delta\tau}. \label{eq:phi_imag_def}
\end{align}
These phase expressions are subsequently used to compute both diagonal and off-diagonal elements of the covariance matrix $\mathbf{R_H}$ through integration over the \gls{wssus} scattering distribution. For diagonal elements ($p=p'$), only $\phi_{\text{real}}$ contributes since $|\exp(\phi)|^2 = \exp(2\phi_{\text{real}})$, proving power invariance to $\alpha_c, \beta_c$. For off-diagonal elements ($p \neq p'$), both $\phi_{\text{real}}$ and $\phi_{\text{imag}}$ enter the complex phase $\Phi(\tau_1, \nu_1)$ in~\eqref{eq:quadratic_phase_general}, enabling closed-form solutions via the complex error function.

\vspace{-1em}
\subsection{The Diagonal Case: A Study of Power Invariance}

We begin our investigation by analyzing the diagonal elements of the covariance matrix, $[\mathbf{R_H}]_{p,p}$, where $p=p'$. These terms are of particular importance as they represent the total average power received at a specific delay-Doppler grid point $p=(k,l)$. A crucial analytical simplification occurs in this case, as the expectation is performed on the squared magnitude of the channel coefficient, $|H_{p,q}|^2$. The squared magnitude of the \gls{tgp} shaping term is
\begin{align}
    \left| \exp(\phi_{\text{real}} + j\phi_{\text{imag}}) \right|^2 &= \exp(2\phi_{\text{real}}) \nonumber  \\
    &= \exp\left(-\frac{4\pi\Delta\tau^2}{\gamma T^2} - \frac{4\pi\gamma\Delta\nu^2}{B^2} \right),
\end{align}
where $\Delta\tau = (k-m)\Delta\tau - \tau_1$ and $\Delta\nu = (l-n)\Delta\nu - \nu_1$. The phase-warping term $\phi_{\text{imag}}$, which contains $\alpha_c$ and $\beta_c$, is completely eliminated. Consequently, the the expected squared magnitude is fundamentally independent of $\alpha_c$ and $\beta_c$. Next, performing the 2D integral over the \gls{wssus} channel statistics yields a closed-form expression for the diagonal elements
\begin{equation}
    [\mathbf{R_H}]_{p,p} = \frac{P\sigma_h^2}{2\tau_{\max}\nu_{\max}} \sum_{m=0}^{M-1}\sum_{n=0}^{N-1} \mathcal{I}_\tau(k,m) \cdot \mathcal{I}_\nu(l,n),
    \label{eq:rh_diag_final}
\end{equation}
where $\mathcal{I}_\tau$ and $\mathcal{I}_\nu$ are the integrals of the truncated Gaussian envelope. Their solutions are given by:
\begin{align}
    \mathcal{I}_\tau(k,m) &= \frac{\sqrt{\pi}}{2\sqrt{A_\tau}}\left[\erf(\zeta_{\tau,u})-\erf(\zeta_{\tau,l})\right], \label{eq:I_tau_diag} \\
    \mathcal{I}_\nu(l,n) &= \frac{\sqrt{\pi}}{2\sqrt{A_\nu}}\left[\erf(\zeta_{\nu,u})-\erf(\zeta_{\nu,l})\right], \label{eq:I_nu_diag}
\end{align}
where error function, $\erf(\cdot)$, arguments are defined as $\zeta_{\tau,u} = \sqrt{A_\tau}(\tau_{\max}-c_\tau)$ and $\zeta_{\tau,l} = -\sqrt{A_\tau}c_\tau$ for the delay integral, and as $\zeta_{\nu,u} = \sqrt{A_\nu}(\nu_{\max}-c_\nu)$ and $\zeta_{\nu,l} = -\sqrt{A_\nu}(\nu_{\max}+c_\nu)$ for the Doppler integral. The remaining constants are given by $c_\tau=(k-m)\Delta\tau$, $c_\nu=(l-n)\Delta\nu$, $A_\tau=4\pi/(\gamma T^2)$, and $A_\nu=4\pi\gamma/B^2$. The expressions in \eqref{eq:rh_diag_final}--\eqref{eq:I_nu_diag} provide a profound insight: the average power at any \gls{dd} grid point depends only on the \gls{tgp}'s envelope-shaping parameter, $\gamma$, and is invariant to the phase-warping parameters, $\alpha_c$ and $\beta_c$. This finding is a cornerstone of our \gls{isac} trade-off analysis. It reveals that any  capacity loss is not due to a reduction in average per-bin received power but from changes in the \emph{structure} of the covariance matrix, specifically, the introduction of deleterious correlations between distinct delay–Doppler bins captured by the off-diagonal elements.

\vspace{-0.5em}
\subsection{The Off-Diagonal Case: The Interference Structure}

To resolve the paradox presented by the power invariance of the diagonal elements, we now turn to the general, off-diagonal terms of the covariance matrix, $[\mathbf{R_H}]_{p,p'}$, where $p \neq p'$. These terms quantify the statistical correlation, or average interference, between distinct delay-Doppler grid points. Unlike the diagonal case, the analysis involves the product of two different complex channel coefficients, $H_{p,q}$ and $H_{p',q}^*$, whose phases do not cancel out.

As established in our analytical framework, the total complex exponent of the integrand, $\Phi(\tau_1, \nu_1)$, simplifies to a separable quadratic polynomial:
\begin{equation}
    \Phi(\tau_1, \nu_1) = (A_\tau \tau_1^2 + B_\tau \tau_1) + (A_\nu \nu_1^2 + B_\nu \nu_1) + C_0.
    \label{eq:quadratic_phase_general}
\end{equation}
Crucially, for the off-diagonal case, the coefficients $\{A_\tau, B_\tau, A_\nu, B_\nu, C_0\}$ are complex-valued and depend on the full set of \gls{tgp} parameters, including the phase-warping terms $\alpha_c$ and $\beta_c$. Their complete expressions are detailed in Appendix~\ref{app:covariance_coeffs}. This separability once again allows the 2D expectation integral to be factored into two 1D complex Gaussian integrals, $\mathcal{I}_\tau$ and $\mathcal{I}_\nu$. The closed-form expression for the off-diagonal correlation is
\begin{equation}
    \mathbb{E}\left[H_{p,q}^{(1)} \left(H_{p',q}^{(1)}\right)^*\right] = \frac{\sigma_h^2}{2\tau_{\max}\nu_{\max}} e^{C_0} \cdot \mathcal{I}_\tau \cdot \mathcal{I}_\nu,
    \label{eq:covariance_component_general}
\end{equation}
where the solutions to the complex integrals are given in terms of the complex error function:
\begin{align}
    \mathcal{I}_\tau &= \frac{1}{2}\sqrt{\frac{\pi}{-A_\tau}} e^{-\frac{B_\tau^2}{4A_\tau}} \left[ \erf(\xi_{\tau,u}) - \erf(\xi_{\tau,l}) \right], \label{eq:I_tau_general} \\
    \mathcal{I}_\nu &= \frac{1}{2}\sqrt{\frac{\pi}{-A_\nu}} e^{-\frac{B_\nu^2}{4A_\nu}} \left[ \erf(\xi_{\nu,u}) - \erf(\xi_{\nu,l}) \right]. \label{eq:I_nu_general}
\end{align}
To maintain clarity, the complex arguments of the error functions are parametrized as
\begin{align}
    \xi_{\tau,u} &= \frac{2A_\tau\tau_{\max}+B_\tau}{2\sqrt{-A_\tau}}, & \xi_{\tau,l} &= \frac{B_\tau}{2\sqrt{-A_\tau}}, \label{eq:xi_tau_def}\\
    \xi_{\nu,u} &= \frac{2A_\nu\nu_{\max}+B_\nu}{2\sqrt{-A_\nu}}, & \xi_{\nu,l} &= \frac{-2A_\nu\nu_{\max}+B_\nu}{2\sqrt{-A_\nu}}. \label{eq:xi_nu_def}
\end{align}
This result decisively resolves the paradox. While $\gamma$ governs the magnitude of the interference footprint, the phase-warping parameters $(\alpha_c,\beta_c)$ are the master controls for the channel's complex correlation structure. By precisely manipulating phase relationships between symbols, these parameters deterministically reconfigure the statistical correlations across the entire delay-Doppler grid, directly altering the singular value spectrum of $\mathbf{R_H}$ and, consequently, its determinant. This establishes the analytical link between the \gls{tgp}'s phase parameters and communication performance: sensing-oriented pulse design via $(\alpha_c,\beta_c)$ directly impacts capacity through the covariance structure $\mathbf{R_H}=\mathbb{E}[\mathbf{H}\mathbf{H}^H]$ in \eqref{eq:jensen_bound}.

\vspace{-1em}
\subsection{Numerical Validation and Performance Analysis}
\label{subsec:numerical_validation}

This section provides numerical insights into the \gls{tgp}'s performance mechanisms. Building on our analytical framework, we investigate how parameters ($\alpha_c, \beta_c$) fundamentally reshape the effective channel. We employ three complementary metrics: (i) Jensen capacity as the primary performance indicator; (ii) the \gls{ipr}, defined as
\begin{equation}
\text{IPR} \triangleq \frac{\sum_{j\neq k}|[\mathbf{R_H}]_{j,k}|^2}{\sum_{j}|[\mathbf{R_H}]_{j,j}|^2},
\end{equation}
quantifying the relative strength of inter-symbol interference; and (iii) the condition number $\kappa(\mathbf{R_H}) = \lambda_{\max}/\lambda_{\min}$, a key metric for evaluating channel conditioning and numerical stability~\cite{Heath2005}. Together, these metrics expose the underlying interference mitigation mechanisms. Next, to analyze these effects, we adopt a canonical \gls{wssus} channel model with $P=8$ paths, assumed to be independently and uniformly distributed within the delay and Doppler spreads ($\tau_{\max}$, $\nu_{\max}$)~\cite{Jung2007}. The parameters are specified in Table~\ref{tab:sim_params}.

\begin{table}[!t]
\renewcommand{\arraystretch}{1.2}
\caption{Simulation Parameters for the Covariance Analysis.}
\label{tab:sim_params}
\centering
\begin{tabular}{ll}
\toprule
\textbf{Parameter} & \textbf{Value} \\
\midrule
\gls{dd} Grid Size ($M \times N$) & $8 \times 8$ \\
Frame Duration ($T$) & $1.12$ ms \\
Signal Bandwidth ($B$) & $28$ kHz \\
Number of Channel Paths ($P$) & $8$ \\
Maximum Delay Spread ($\tau_{\max}$) & $5 \times (T/M)$ \\
Maximum Doppler Spread ($\nu_{\max}$) & $2 \times (B/N)$ \\
Signal-to-Noise Ratio (SNR) & $15$ dB (unless specified) \\
Monte Carlo Simulations & 1000 \\
\bottomrule
\end{tabular}
\end{table}

\vspace{0.1cm}
\subsubsection{Quantifying the Impact of \gls{tgp} Phase Warping}

\begin{figure}[!t]
    \centering
    \begin{subfigure}{0.48\columnwidth}
        \centering
        \includegraphics[width=0.95\textwidth]{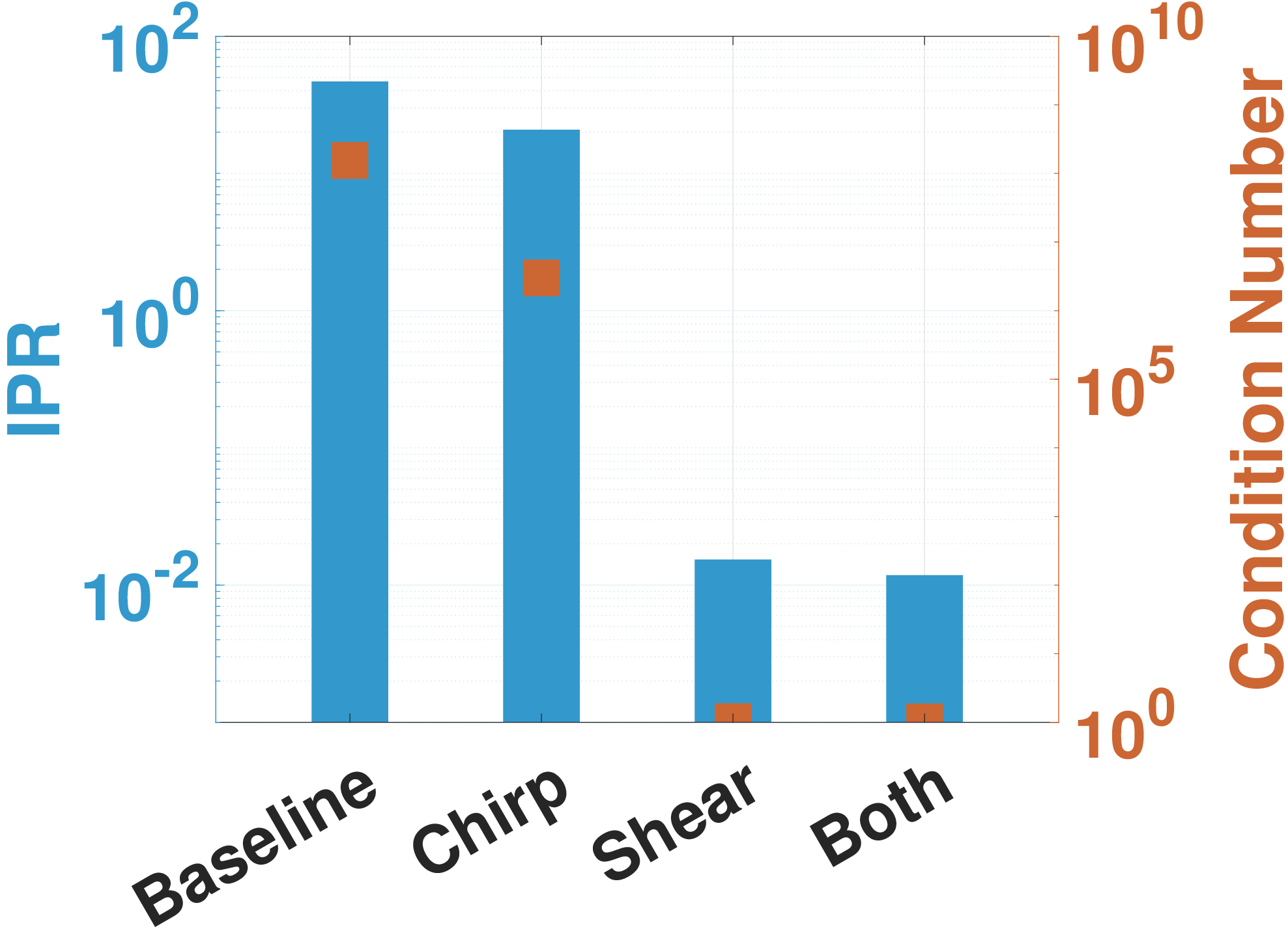}
        \caption{Channel Conditioning.}
        \label{fig:ipr_cond}
    \end{subfigure}
    \vspace{0.5cm}    
    \begin{subfigure}{0.48\columnwidth}
        \centering
        \includegraphics[width=0.95\textwidth]{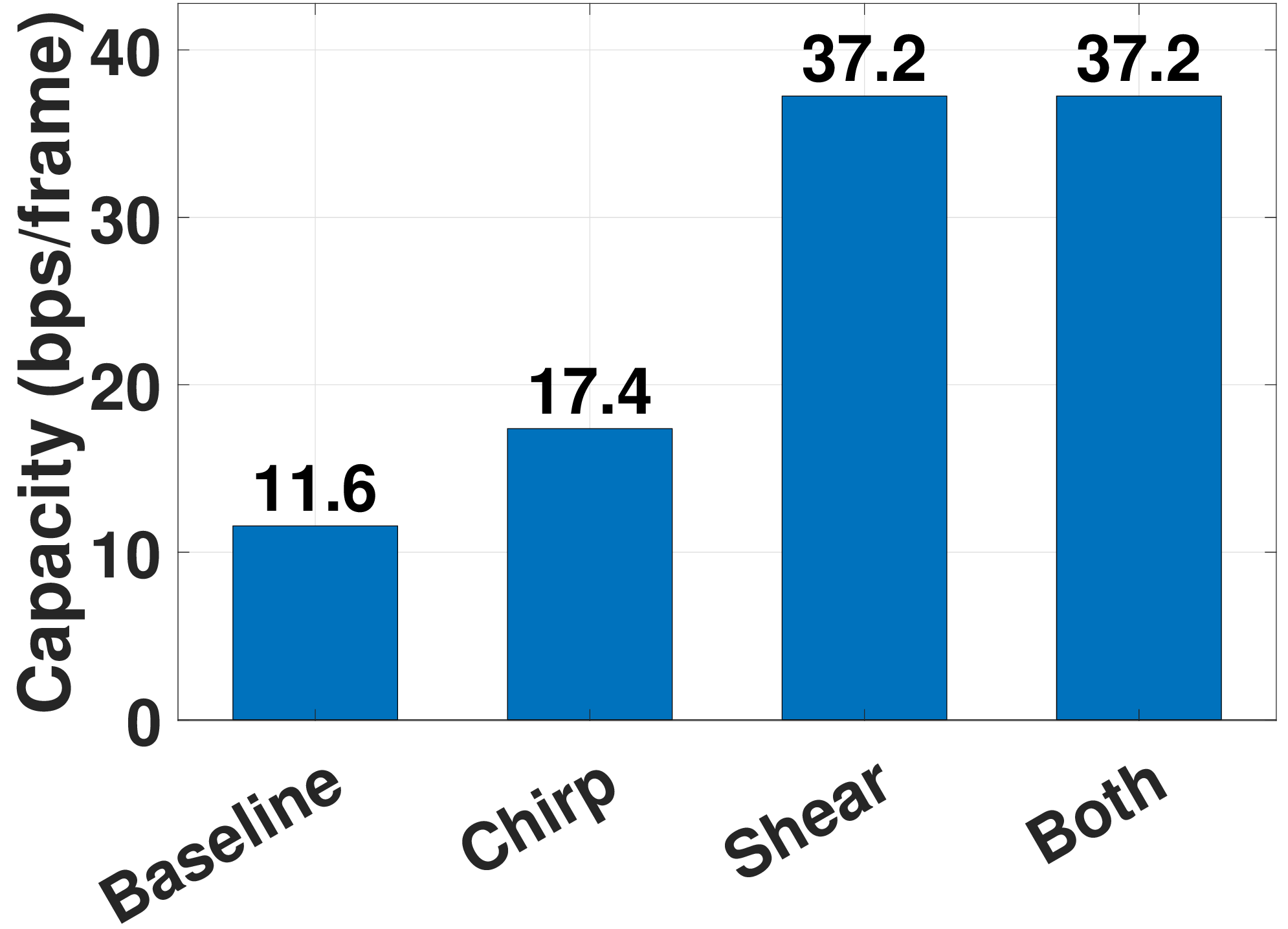}
        \caption{Resulting Capacity Gain.}
        \label{fig:capacity_gain}
    \end{subfigure}
    \vspace{-1em}
    \caption{\small Impact of \gls{tgp} phase-warping. (a) Channel structure improvement via shear parameter ($\beta_c$). (b) Resulting gain in communication capacity.}
    \vspace{-2em}
    \label{fig:tgp_impact_vertical}
\end{figure}

We analyze the phase-warping mechanisms by comparing four configurations: Baseline \gls{sgp} \cite{Das2025a}  ($\alpha_c=\beta_c=0$), Chirp-only \cite{Li2025a} ($\alpha_c=5$), Shear-only \cite{Ozaktas2001} ($\beta_c=5$), and Combined ($\alpha_c=\beta_c=5$). Fig.~\ref{fig:tgp_impact_vertical} reveals a stark performance hierarchy. The unmodulated \gls{sgp} suffers from severe ill-conditioning (condition number $\approx 1.5 \times 10^8$) and high \gls{ipr}, capping capacity at 11.6~bps/frame. While $\alpha_c$ provides moderate gains, activating $\beta_c$ fundamentally restructures the channel. Consistent with the mechanism proven in Section~\ref{sec:performance_analysis}, the shear parameter near-diagonalizes the effective channel matrix, dropping the \gls{ipr} to 0.015 and the condition number to a near-ideal 1.02. This structural improvement triples the capacity to 37.3~bps/frame. Notably, adding chirp on top of shear yields negligible further benefit, confirming $\beta_c$ as the dominant \gls{isi} mitigation mechanism.

\section{Sensing Performance Analysis}
\label{sec:sensing_performance}

We assess the sensing capabilities of the proposed \gls{tgp} via two complementary tools: \gls{af}, which provides a qualitative picture of resolution and sidelobes, and \gls{fim}, which quantifies fundamental estimation precision.
\paragraph{\gls{af}} Since we adopt the twisted-convolution convention used in Section~\ref{subsec:sensing_model}, the \gls{af} of the \gls{dd}-domain pulse $x_{\mathrm{dd}}$ is
\begin{equation}
\begin{aligned}
AF(\tau,\nu)\triangleq \iint &\, x_{\mathrm{dd}}(u,v)\,x_{\mathrm{dd}}^{*}(u-\tau,v-\nu)\\
&\times e^{j2\pi(\nu u-\tau v)}\,du\,dv.
\end{aligned}
\label{eq:af_def}
\end{equation}
Following \eqref{eq:af_def}, two principal cuts summarize resolution:
\begin{itemize}
    \item \textbf{Zero-Doppler cut} $AF(\tau,0)$: equals the \gls{acf} in delay; its mainlobe width sets delay (range) resolution and its sidelobes govern masking of weak targets.
    \item \textbf{Zero-delay cut} $AF(0,\nu)$: Doppler \gls{acf}; its mainlobe width sets velocity resolution.
\end{itemize}
\paragraph{\gls{fim}}
The \gls{fim} at the true target state $\bm{\psi}=[\tau_T,\nu_T]^T$ depends on the \gls{af}’s local curvature at the origin. With the continuous \gls{dd}-noise model in \eqref{eq:dd_noise_psd} and total transmit energy $E_s$, the \gls{fim} can be written as
\begin{equation}\label{eq:fim_def}
\mathbf{I}(\bm{\psi}) \;=\; K_{\mathrm{FIM}}\,
\begin{bmatrix}
I_{\tau\tau}^{(0)} & I_{\tau\nu}^{(0)} \\
I_{\tau\nu}^{(0)} & I_{\nu\nu}^{(0)}
\end{bmatrix},
\,
K_{\mathrm{FIM}} \triangleq \frac{2E_s}{N_0}|h_T|^2,
\end{equation}
where $\{I_{\tau\tau}^{(0)},I_{\nu\nu}^{(0)},I_{\tau\nu}^{(0)}\}$ depend solely on the \gls{dd} pulse. To ensure consistency with Section~\ref{subsec:sensing_model}, we use a \emph{unit-energy} \gls{dd} pulse, $\iint |x_{\mathrm{dd}}|^2=1$, and collect $E_s$ into $K_{\mathrm{FIM}}$. If one instead starts from the unnormalized Gaussian of \eqref{eq:tgp_waveform_explicit} with energy\footnote{This result is obtained by solving the 2D Gaussian integral $\int e^{-a\tau^2}d\tau \cdot \int e^{-b\nu^2}d\nu$ where $a = 4\pi/(\gamma T^2)$ and $b = 4\pi\gamma/B^2$, yielding $(\sqrt{\pi/a})(\sqrt{\pi/b}) = (T\sqrt{\gamma}/2)(B/(2\sqrt{\gamma})) = BT/4$.} $E_0=\iint|x_{\mathrm{dd}}|^2\,d\tau\,d\nu=BT/4$, the unit-energy pulse is $x_{\mathrm{dd}}/\sqrt{E_0}$ and all \gls{fim} entries scale by $1/E_0$ (hence $K_{\mathrm{FIM}}$ would be multiplied by $E_0$). The closed forms are stated in the following for the unit-energy convention.

\par With \gls{fim} defined in \eqref{eq:fim_def}, a key contribution of this work is the analytical derivation of its intrinsic elements for the \gls{tgp}. As derived step-by-step in Appendix~\ref{app:fim_derivation}, these expressions for the unit-energy pulse are given by: 
\begin{align}
    I_{\tau\tau}^{(0)} &= \frac{\pi}{8\gamma T^2} \Big( 4(4+\alpha_{\gamma}^2) + B_{T}^2 (\beta_c-2)^2 \Big), \label{eq:Itau} \\
    I_{\nu\nu}^{(0)} &= \frac{\pi\gamma}{8B^2} \Big( 16 + B_{T}^2(\beta_c+2)^2 \Big), \label{eq:Inu} \\
    I_{\tau\nu}^{(0)} &= \frac{\pi}{4}\alpha_c \gamma(2+\beta_c), \label{eq:Icross}
\end{align}
where $B_{T}^2=B^2 T^2$ and $\alpha_{\gamma}^2=\alpha_c^2\gamma^2$. Hence $\gamma$ trades axial curvatures (delay vs. Doppler), while $\alpha_c$ and $\beta_c$ steer the mixed curvature.
\paragraph{\glspl{crlb} and Performance Trade-offs}
The \gls{crlb} on the estimation of $(\tau_T, \nu_T)$ is given by the diagonal elements of the inverse \gls{fim}, $\mathbf{I}^{-1}$. By substituting the intrinsic elements from \eqref{eq:Itau}--\eqref{eq:Icross}, we obtain the complete closed-form expressions for the bounds of delay and Dopper as:
\begin{align}
    \mathrm{CRLB}(\tau_T) &= \frac{8 T^2 \gamma (16+B_{T}^2 (\beta_c+2)^2)}{K_{\mathrm{FIM}} \pi \cdot D_{\mathrm{poly}}(\alpha_c, \beta_c, \gamma)}, \label{eq:crlb_tau_final} \\ \nonumber \\ 
    \mathrm{CRLB}(\nu_T) &= \frac{8 B^2 (16+B_{T}^2(\beta_c-2)^2+4 \alpha_{\gamma}^2)}{K_{\mathrm{FIM}} \pi \gamma \cdot D_{\mathrm{poly}}(\alpha_c, \beta_c, \gamma)}, \label{eq:crlb_nu_final}
\end{align}
where the denominator polynomial, $D_{\mathrm{poly}}(\cdot)$, in \eqref{eq:crlb_tau_final} and \eqref{eq:crlb_nu_final} is defined as
\begin{align}
    D_{\mathrm{poly}} = &B_{T}^4(\beta_c^2-4)^2 + 32 B_{T}^2(\beta_c^2+4) + 64 (4+\alpha_{\gamma}^2).
\end{align}
To better understand the structure of these bounds, it is insightful to express them in their more fundamental form, which isolates the effect of parameter coupling. This is achieved using the estimator correlation coefficient, $\rho$:
\begin{align}
    \mathrm{\gls{crlb}}(\tau_T) &= \frac{1}{K_{\mathrm{FIM}}\,I_{\tau\tau}^{(0)}} \,\frac{1}{1-\rho^2}, \label{eq:crb_tau}\\
    \mathrm{\gls{crlb}}(\nu_T)  &= \frac{1}{K_{\mathrm{FIM}}\,I_{\nu\nu}^{(0)}} \,\frac{1}{1-\rho^2}, \label{eq:crb_nu}
\end{align}
where the term $(1-\rho^2)^{-1} \ge 1$ is the "coupling penalty." The squared correlation coefficient, $\rho^2 \triangleq (I_{\tau\nu}^{(0)})^2/(I_{\tau\tau}^{(0)}I_{\nu\nu}^{(0)})$, is given by:
\begin{equation}
\rho^2
=\frac{
4 B_{T}^2(2+\beta_c)^2 \alpha_{\gamma}^2
}{(16+B_{T}^2(2+\beta_c)^2)
(B_{T}^2(\beta_c-2)^2 + 16+4\alpha_{\gamma}^2)
},
\label{eq:rho2}
\end{equation}
From \eqref{eq:rho2}, we deduce that the estimator coupling can be completely eliminated ($\rho^2 = 0$) if either the chirp is deactivated ($\alpha_c=0$) or the phase-coupling is set to $\beta_c=-2$.

\begin{figure}[!t]
    \centering
    \includegraphics[width=0.75\columnwidth]{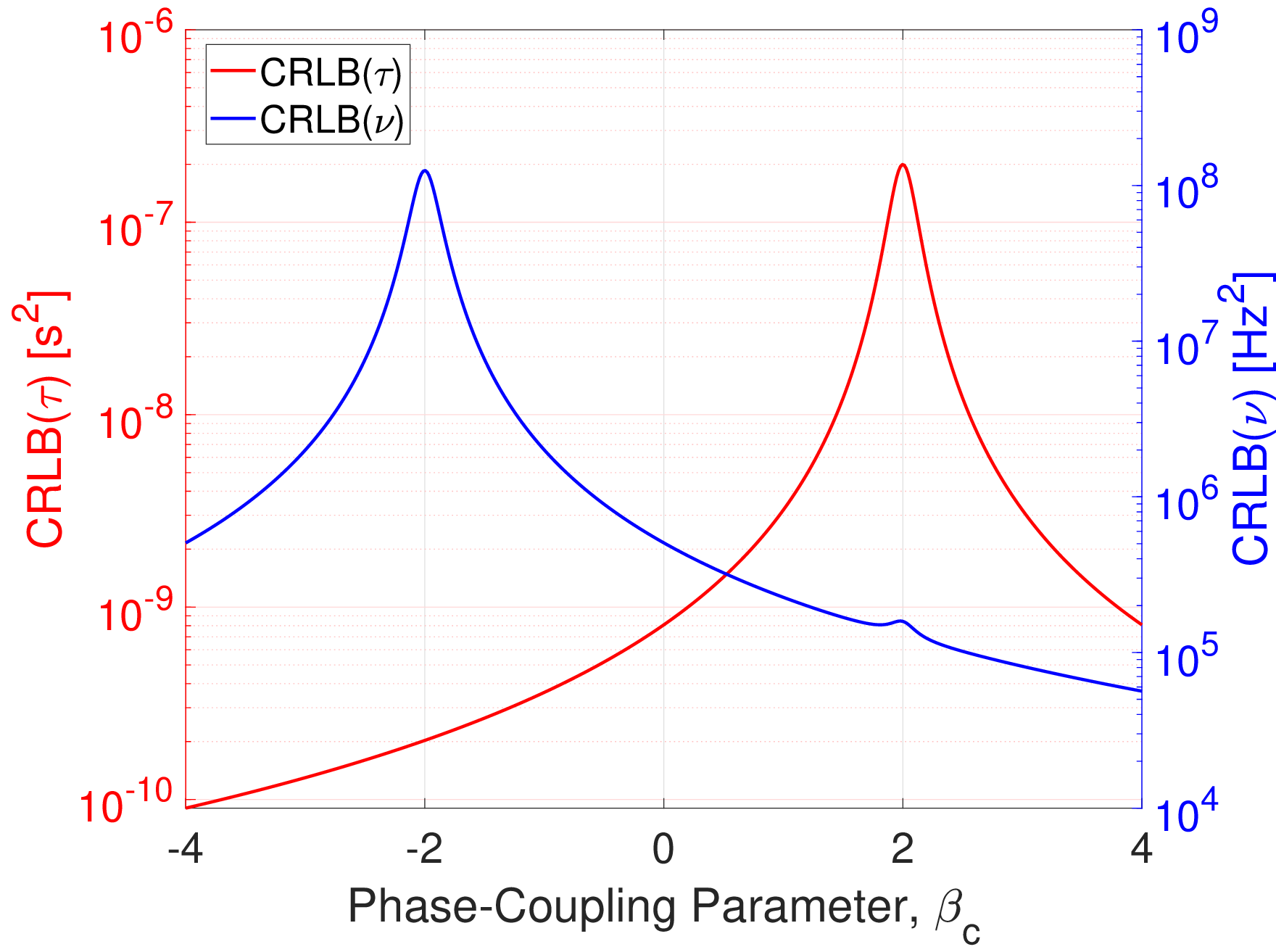}
    \caption{\small CRLB for delay and Doppler as a function of the phase-coupling parameter $\beta_c$.}
    \label{fig:crlb_vs_beta}
    \vspace{-1em}
\end{figure}

While these expressions are complex, their structure reveals a clear and actionable trade-off, which is illustrated in Fig.~\ref{fig:crlb_vs_beta} for our system design ($B=28$~kHz, $T=1.12$~ms). We focus on the impact of $\beta_c$, as its dominant, quartic influence on performance is evident from its role in the $D_{\mathrm{poly}}$ term. The plot, which sweeps $\beta_c$ from $-4$ to $4$ while keeping $\gamma$ and $\alpha_c$ fixed, uses a dual-axis scale to demonstrate a stark inverse relationship: the choice of $\beta_c$ that optimizes delay precision simultaneously degrades Doppler precision, and vice-versa.

The analysis of the plot reveals several key behaviors. The worst-case performance for each metric occurs near the point that is optimal for the other; specifically, the peak of $\mathrm{CRLB}(\tau_T)$ occurs near $\beta_c=2$, while the peak of $\mathrm{CRLB}(\nu_T)$ is found near $\beta_c=-2$. Conversely, optimal performance is achieved as the magnitude of $|\beta_c|$ increases, which grows the denominator $D_{\mathrm{poly}}$ and drives both bounds down. For the balanced case shown, at the edge of the observed range, the minimum for delay precision reaches $\mathrm{CRLB}(\tau_T) \approx 10^{-10}\,\mathrm{s}^2$ at $\beta_c = -4$. This bound corresponds to a timing error standard deviation of only $\sigma_\tau \approx 10~\mu\text{s}$. Inversely, the minimum for Doppler precision is achieved at the symmetric point $\beta_c=4$, reaching a value of $\mathrm{CRLB}(\nu_T) \approx 5 \times 10^4\,\mathrm{Hz}^2$. This corresponds to a velocity estimation standard deviation of $\sigma_\nu \approx 224~\text{Hz}$, demonstrating the \gls{tgp}'s capability to achieve high precision in either domain via careful tuning.
\vspace{0.1cm}
\paragraph{Scalar Sensing Figures of Merit}
To capture the joint estimability of the target with a single metric, we use the \gls{fim} determinant, $Q_{\det} \triangleq \det \mathbf{I}(\bm{\psi})$. Substituting the derived \gls{fim} components from \eqref{eq:Itau}--\eqref{eq:Icross} and simplifying yields the compact closed-form expression:
\begin{align}
    Q_{\det} = \frac{\pi^2 K_{\mathrm{FIM}}^2}{64 B^2 T^2}D_{\mathrm{poly}}(\alpha_c, \beta_c, \gamma).
    \label{eq:det_final}
\end{align}
This result provides critical insights. As $Q_{\det}$ is inversely proportional to the area of the \gls{crlb} uncertainty ellipse, a larger value signifies higher joint precision \cite{Xinyang2023}. 
\section{\gls{isac} Numerical Results}
\label{sec:numerical_results}
%
This section validates the analysis and demonstrates the performance of the proposed \gls{dd}-domain \gls{tgp} via a comprehensive simulation study. To rigorously evaluate the \gls{isac} trade-off, our methodology is structured into two complementary experiments: first, a communication-centric analysis using ergodic capacity, and second, a sensing-centric analysis using the fundamental \gls{crlb}. To assess performance in realistic conditions, we employ the 3GPP Vehicular-A (Veh-A) six-path channel model, a widely-used high-mobility benchmark in pulse-shaping literature~\cite{Mehrotra2025,Jesbin2025,Das2025a,ITUM1225}.
The Doppler shift of the $i$-th path is $\nu_i = \nu_{\max} \cos \theta_i$, where $\nu_{\max} = 815$~Hz and $\theta_i \sim \mathcal{U}[0, 2\pi)$ \cite{Jesbin2025}. This standardized model considers non-uniform scattering, moving beyond the \gls{wssus} channel model used in Section \ref{subsec:numerical_validation}. The complete parameters for both experimental setups are explicitly detailed in Table~\ref{tab:sim_parameters}. 

We first present the performance of the \gls{tgp} against standard \gls{dd} benchmarks (\gls{sgp}, \gls{rrc}, and Sinc) within each domain. We then combine these results to present the achievable \gls{isac} trade-off curves, demonstrating how the \gls{tgp} parameters $(\gamma,\alpha_c,\beta_c)$ navigate the Pareto frontier. 

\begin{table}[!h]
\vspace{-0.2em}
\centering
\caption{Simulation Parameters for \gls{isac} Analysis}
\vspace{-1em}
\label{tab:sim_parameters}
\renewcommand{\arraystretch}{1.1} 
\begin{tabular}{ll}
\toprule
\textbf{Parameter} & \textbf{Value} \\
\midrule
\multicolumn{2}{l}{\textit{\textbf{Fundamental Physical Parameters (Common)}}} \\
\quad Frame Duration ($T$) & 1.12 ms \\
\quad Bandwidth ($B$) & 28 kHz \\
\quad \gls{snr} Range & 0 to 20 dB (4 dB steps) \\
\quad Benchmark Pulses & Gaussian (\gls{sgp}), \gls{rrc}, Sinc \\
\quad \gls{rrc} Roll-off Factor ($\beta_{\text{RRC}}$) & 0.6 \\
\midrule
\multicolumn{2}{l}{\textit{\textbf{\gls{tgp} Parameter Sweep (Common)}}} \\
\quad Sweep Resolution ($n$) & 20 \\
\quad Total \gls{tgp} Points & $20^3 = 8,000$ \\
\quad Aspect Ratio ($\gamma$) & $\gamma \in [0.01, 100]$ (log scale) \\
\quad Chirp Rate ($\alpha_c$) & $\alpha_c \in [0, 50]$ (linear scale) \\
\quad Phase Coupling ($\beta_c$) & $\beta_c \in [0, 10]$ (linear scale) \\
\midrule
\multicolumn{2}{l}{\textit{\textbf{Communication (Capacity) Analysis Parameters}}} \\
\quad Metric & Ergodic Capacity $\mathbb{E}[C(H)]$ \\
\quad \gls{dd} Grid ($M \times N$) & 16 $\times$ 16 \\
\quad Evaluation Method & $\#$ Average \\
\quad $\#$ Realizations & 160 \\
\quad Channel Model & 3GPP Vehicular-A (6 paths) \cite{ITUM1225}\\
\quad Maximum Doppler ($\nu_{\max}$) & 815 Hz \cite{Mehrotra2025,Jesbin2025,Das2025a}\\
\quad Doppler Distribution & $\nu_i = \nu_{\max} \cos \theta_i$, $\theta_i \sim \mathcal{U}[0, 2\pi)$\\
\midrule
\multicolumn{2}{l}{\textit{\textbf{Sensing (\gls{crlb}) Analysis Parameters}}} \\
\quad Metric & \gls{crlb} \\
\quad \gls{dd} Grid ($M \times N$) & 256 $\times$ 256 \\
\quad Evaluation Method & Numerical (Discrete \gls{fim}) \\
\bottomrule
\end{tabular}
\vspace{-0.5em}
\end{table}

\begin{figure*}[!h]
\centering   
\begin{subfigure}{0.48\textwidth}
\centering
\includegraphics[width=0.95\textwidth]{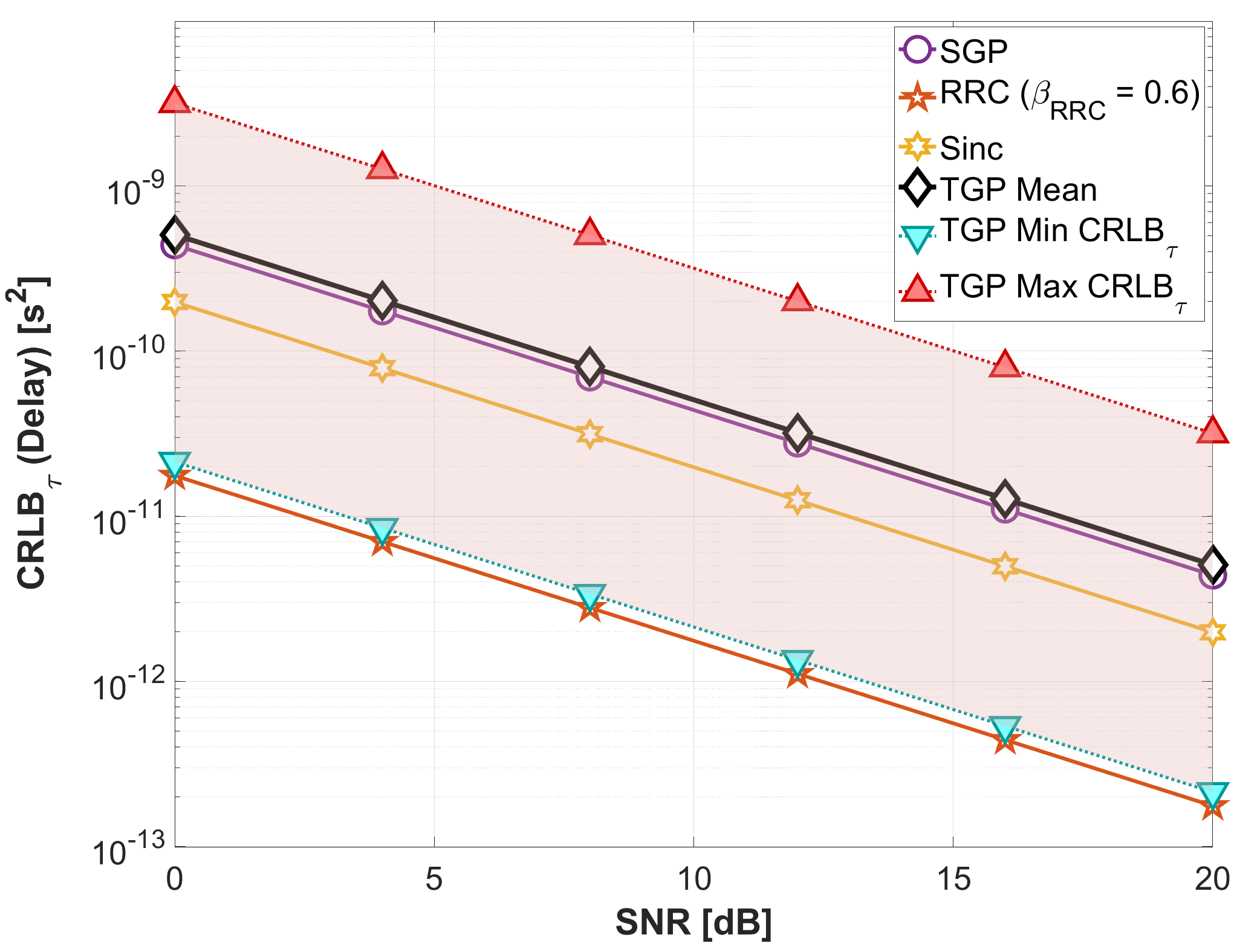}
        \caption{Sensing Performance (Delay \gls{crlb} vs. \gls{snr})}
        \label{fig:CRBtauSNR}
    \end{subfigure}
    \vspace{0.5cm}   
    \begin{subfigure}{0.48\textwidth}
        \centering
        \includegraphics[width=0.95\textwidth]{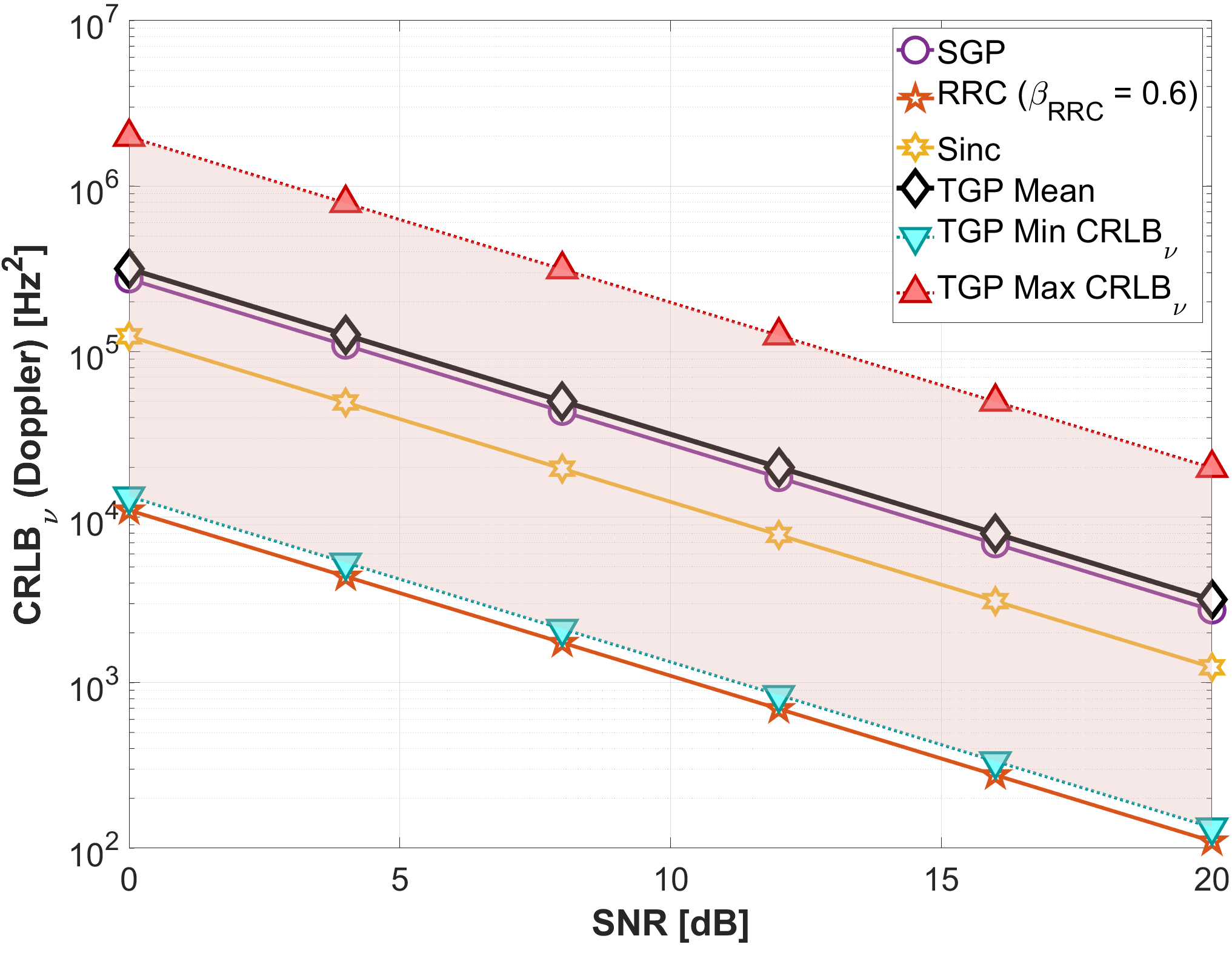}
        \caption{Sensing Performance (Doppler \gls{crlb} vs. \gls{snr})}
        \label{fig:CRBnuSNR}
    \end{subfigure}
    \vspace{-1em}
    \caption{\small Evaluation of sensing performance versus \gls{snr}, based on the parameters in Table~\ref{tab:sim_parameters}, the shaded region
illustrates the achievable TGP envelope. (a) Delay \gls{crlb} performance of benchmarks. (b) Doppler \gls{crlb} performance of benchmarks.}
\vspace{-1em}
    \label{fig:sensing_performance_vs_snr} 
\end{figure*}

\subsection{Practical Parameter Constraints and Operating Ranges}
\label{sec:practical_limitations}

While the parametric analysis in Sections~\ref{sec:performance_analysis}  and \ref{sec:sensing_performance} reveals the theoretical benefits of the \gls{tgp}, practical deployments are bounded by fundamental trade-offs in energy containment, receiver complexity, and hardware robustness. For our \gls{dd} grid configuration (e.g., $M=16, N=16$, $T=1.12$~ms, $B=28$~kHz), the parameter ranges explored in our numerical analysis are justified as follows:

\paragraph{Aspect-Ratio ($\gamma \in [0.01, 100]$)}
This parameter is fundamentally limited by energy containment, as dictated by the Balian-Low Theorem~\cite{Jesbin2025}. The pulse envelope must be sufficiently localized within the $T \times B$ block to preserve the quasi-interference-free "crystalline region"~\cite{Gopalam2024}.
\begin{itemize}
    \item A choice of $\gamma \gg 1$ expands the pulse in the delay domain ($\sigma_{\tau}$ increases), causing energy to leak into adjacent time blocks, which manifests as \gls{isi}.
    \item Conversely, $\gamma \ll 1$ expands the pulse in the Doppler domain ($\sigma_{\nu}$ increases), causing energy to leak into adjacent frequency bins, which manifests as \gls{ici}~\cite{Jesbin2025}.
\end{itemize}
The sweep range $\gamma \in [0.01, 100]$ is thus chosen to explore this entire practical trade-off, from delay-dominated to Doppler-dominated pulse geometries.

\paragraph{Chirp-Rate ($\alpha_c \in [0, 50]$)}
This parameter is limited by a direct conflict between \gls{isac} requirements. On one hand, sensing precision benefits from $\alpha_c > 0$ to induce the critical delay-Doppler coupling $I_{\tau\nu}$ in the \gls{fim}, as shown in our analysis. On the other hand, practical, low-complexity \gls{mmse} receivers for communication are severely penalized by arbitrary chirp rates. Analogous \gls{afdm} systems, for example, only achieve optimal \gls{ber} for small, discrete chirp parameter sets~\cite{Li2025a}. Our sweep range $\alpha_c \in [0, 50]$ is therefore selected to explore this conflict and identify the operational frontier where sensing optimization begins to catastrophically degrade communication performance.

\paragraph{Phase-Coupling ($\beta_c \in [0, 10]$)}
This parameter is limited not by theoretical capacity, which, as our analysis shows, it often increases, but by phase aliasing~\cite{Mohammed2023} and hardware robustness~\cite{Bjornson2013}. The discrete $M \times N$ grid imposes a Nyquist-like sampling limit on the bilinear phase in \eqref{eq:Phase_Warping_Function}, $\phi(\tau, \nu) = \pi \beta_c \tau \nu$. The maximum phase wrap between adjacent grid samples must remain unambiguous. We generalize this limit to tolerate a specific number of integer phase wraps, $K_{alias} \in \mathbb{Z}_{\ge 0}$. Based on this derivation, the maximum allowable $\beta_c$ for a given $K_{alias}$ is found to be:
\begin{equation}
    \beta_{c, \text{limit}}(K_{alias}) \approx (K_{alias}+1) \cdot \frac{2 \min(M, N)}{BT}
    \label{eq:beta_k_limit}
\end{equation}
The conventional zero-aliasing (Nyquist) condition~\cite{Mohammed2023} corresponds to $K_{alias} = 0$, which for our grid ($M=N=16$, $BT \approx 31.36$) yields $\beta_{c, \text{Nyquist}} \approx 1.028$. However, our numerical results (e.g., Fig.~\ref{fig:capacity_gain}) confirm the system can operate beneficially in a "structured aliasing" regime where $K_{alias} > 0$. The practical upper bound is therefore not this Nyquist limit, but rather hardware robustness~\cite{Bjornson2013}. As $\beta_c$ grows, the phase wraps numerous times between samples, making the receiver hyper-sensitive to synchronization jitter and phase noise~\cite{Surabhi2019}, which can prevent the unambiguous phase resolution required for detection~\cite{Wymeersch2006} and lead to performance collapse. Our chosen range $\beta_c \in [0, 10]$ (corresponding to $K_{alias}$ from 0 to $\approx 9$) thus represents a principled exploration of this robust, practical, and structured aliasing regime.

\subsection{Benchmark Pulses}

To rigorously evaluate the performance of the proposed \gls{tgp}, we compare it against a set of established pulse shapes, including its own baseline. For a fair comparison, all discrete pulse shapes are normalized to have unit energy. In the following, we delineate the pulse shapes considered in our analysis

\paragraph{Proposed: Tunable Gaussian Pulse (\gls{tgp})}
The discrete-domain \gls{tgp}, centered at the origin on an $M \times N$ grid with resolutions $d_\tau=T/M$ and $d_\nu=B/N$, is given by
\begin{align}
x_{\mathrm{dd}}^{\mathrm{TGP}}[m,n] = C \cdot &\exp\!\left(-\pi\left(\frac{2(m d_\tau)^2}{\gamma T^2} + \frac{2\gamma(n d_\nu)^2}{B^2}\right)\right) \nonumber \\
&\hspace{-1cm}\times \exp\!\left(j\pi\left(\frac{\alpha_c}{T^2}(m d_\tau)^2 + \beta_c(m d_\tau)(n d_\nu)\right)\right),
\label{eq:tgp_discrete}
\end{align}
where $m \in \{-M/2, \dots, M/2-1\}$, $n \in \{-N/2, \dots, N/2-1\}$, and $C$ is a normalization constant.
%
\paragraph{Standard Gaussian Pulse (\gls{sgp})}
The \gls{sgp} \cite{Jayachandran2024} is the baseline for comparison and a special case of the \gls{tgp} obtained by setting the tuning parameters to their neutral values: $\gamma=1$, $\alpha_c=0$, and $\beta_c=0$. This simplifies \eqref{eq:tgp_discrete} to
\begin{equation}
x_{\mathrm{dd}}^{\mathrm{SGP}}[m,n] = C \cdot \exp\!\left(-\pi\left(\frac{2(m d_\tau)^2}{T^2} + \frac{2(n d_\nu)^2}{B^2}\right)\right).
\label{eq:sgp_discrete}
\end{equation}
%
\paragraph{Root-Raised Cosine Pulse}
The separable 2D \gls{rrc} pulse \cite{Mohammed2023, mohammed2024otfs, Jayachandran2024, Gopalam2024} is given by
\begin{equation}
    x_{\mathrm{dd}}^{\mathrm{RRC}}[m,n] = C \cdot p_{\mathrm{rrc}}\!\left(\frac{m}{M}, \beta_\tau\right) \cdot p_{\mathrm{rrc}}\!\left(\frac{n(BT)}{N}, \beta_\nu\right),
    \label{eq:rrc_discrete}
\end{equation}
where $C$ is a normalization constant, $p_{\mathrm{rrc}}(\cdot)$ is the standard 1D \gls{rrc} function, and $\beta_\tau, \beta_\nu$ are the delay and Doppler roll-off factors.

\paragraph{Sinc Pulse}
The 2D periodized sinc pulse \cite{Mohammed2023, mohammed2024otfs, Jayachandran2024} is defined by sampling the normalized $\mathrm{sinc}(x) \triangleq \sin(\pi x)/(\pi x)$ function as
\begin{equation}
x_{\mathrm{dd}}^{\mathrm{sinc}}[m,n] = C \cdot \mathrm{sinc}\!\left(m \frac{BT}{M}\right) \cdot \mathrm{sinc}\!\left(n \frac{BT}{N}\right),
\label{eq:sinc_discrete}
\end{equation}
where $C$ is the normalization constant.

\vspace{-0.2em}
\subsection{Sensing Performance Analysis}
\label{subsec:sensing_performance}

We begin our \gls{isac} evaluation by analyzing the sensing performance, with results presented in Fig.~\ref{fig:sensing_performance_vs_snr}. This figure plots the \gls{crlb} versus \gls{snr} for delay estimation (Fig.~\ref{fig:CRBtauSNR}) and Doppler estimation (Fig.~\ref{fig:CRBnuSNR}). We compare the \gls{dd} benchmarks against the complete \gls{tgp} operational envelope, illustrating the best-case (`TGP Min CRLB'), worst-case (`TGP Max CRLB'), and mean (`\gls{tgp} Mean') performance across the $8,000$ configurations from Table~\ref{tab:sim_parameters}. The sensing results in Fig.~\ref{fig:sensing_performance_vs_snr} demonstrate several key characteristics. Both delay and Doppler estimations exhibit similar behavioural hierarchies. The \gls{rrc} benchmark consistently achieves the lowest \gls{crlb}, establishing the performance upper bound for both metrics. Notably, the \gls{tgp} Min \gls{crlb} demonstrates the adaptability of our pulse shaping scheme, approaching the performance of the specialized \gls{rrc} benchmark. At 20 dB, the \gls{tgp}'s minimum delay \gls{crlb} ($2.14\mathrm{e}^{-13}$ s$^2$) reaches within 1.21x of the \gls{rrc} ($1.76\mathrm{e}^{-13}$ s$^2$), with comparable proximity observed for Doppler estimation. Conversely, the Sinc benchmark exhibits poor sensing suitability, degrading performance by a factor of 11.2x relative to \gls{rrc} in both metrics. The \gls{tgp} Mean  and \gls{sgp} demonstrate intermediate, non-competitive performance, while the TGP Max \gls{crlb} confirms that suboptimal \gls{tgp} parameterization results in severe sensing degradation.

\subsection{Communication Performance Analysis}
\label{subsec:comm_performance}

\begin{figure}[!t]
    \centering   
    \includegraphics[width=0.9\columnwidth]{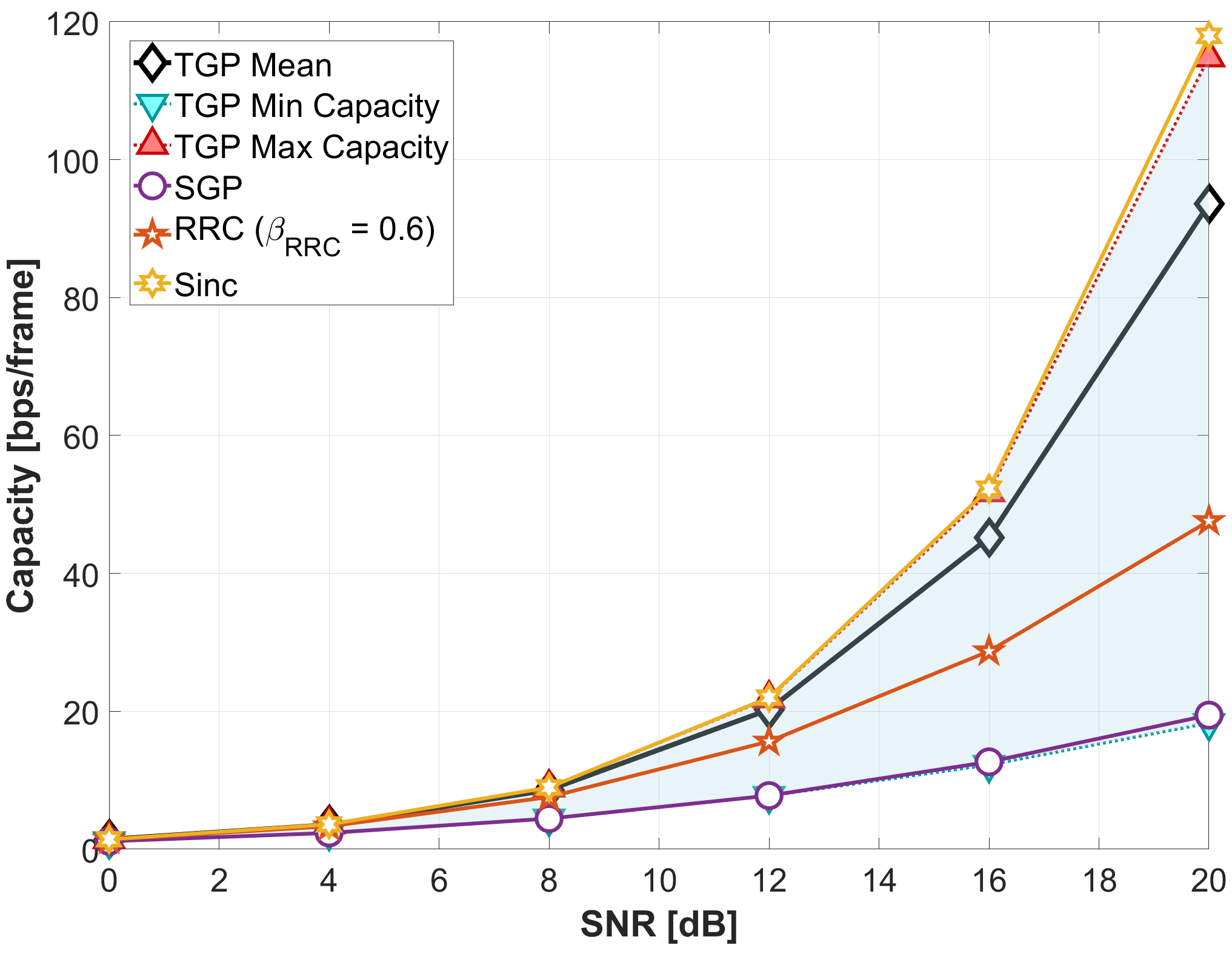}   
    \caption{ \small Capacity vs. \gls{snr}. Shaded region is the\gls{tgp} capacity envelope from parameter sweep (Table~\ref{tab:sim_parameters}).}
    \label{fig:performance_vs_snr} 
    \vspace{-1em}
\end{figure}

We complement the \gls{crlb} analysis by examining communication performance through ergodic capacity evaluation, with results presented in Fig.~\ref{fig:performance_vs_snr}(a). To enable fair comparison of the structural properties induced by different pulse shapes, we normalize each channel realization to unit Frobenius norm ($\|\mathbf{H}\|_F = 1$) before computing \eqref{eq:channel_capacity}. This isolates interference characteristics from absolute power gain variations, distinct from the physical covariance framework in Section~\ref{sub:analytical_framework}. The figure plots capacity versus \gls{snr} for the \gls{dd} benchmarks (Sinc, \gls{rrc}) and the proposed \gls{tgp}. For the \gls{tgp}, we characterize the complete operational envelope derived from the 8,000 parameter configurations detailed in Table~\ref{tab:sim_parameters}, showing the best-case (`\gls{tgp} Max Capacity'), worst-case (`\gls{tgp} Min Capacity'), and mean (`\gls{tgp} Mean') performance.

The results in Fig.~\ref{fig:performance_vs_snr} establish a distinct performance hierarchy. As theoretically anticipated by the \gls{dd} pulse shaping literature~\cite{Das2025a, Jesbin2025, Mehrotra2025}, the Sinc benchmark achieves the highest capacity (117.96~bps/frame at 20~dB). The \gls{rrc} and \gls{sgp} pulses exhibit suboptimal communication performance, reaching only 40.4\% and 16.5\% of the Sinc throughput at 20~dB, respectively. Critically, the \gls{tgp} Max Capacity curve demonstrates the \gls{tgp}'s adaptability, achieving near-optimal performance (114.83~bps/frame at 20~dB) that closely tracks the Sinc benchmark. The extensive shaded region underscores the impact of the tuning parameters $(\gamma,\alpha_c,\beta_c)$, as the performance envelope spans from this near-Sinc maximum down to the \gls{tgp} Min Capacity curve, which exhibits performance comparable to the \gls{sgp} benchmark. Furthermore, the \gls{tgp} Mean curve demonstrates robust average performance, substantially exceeding both \gls{rrc} and \gls{sgp} pulses across the entire \gls{snr} range.
\vspace{-0.3em}
\subsection{\gls{isac} Trade-off Benchmarking Experiment}
\label{subsec:tradeoff_benchmark}

\begin{figure*}[!h]
    \centering   
    \begin{subfigure}{0.48\textwidth}
        \centering
        \includegraphics[width=0.95\textwidth]{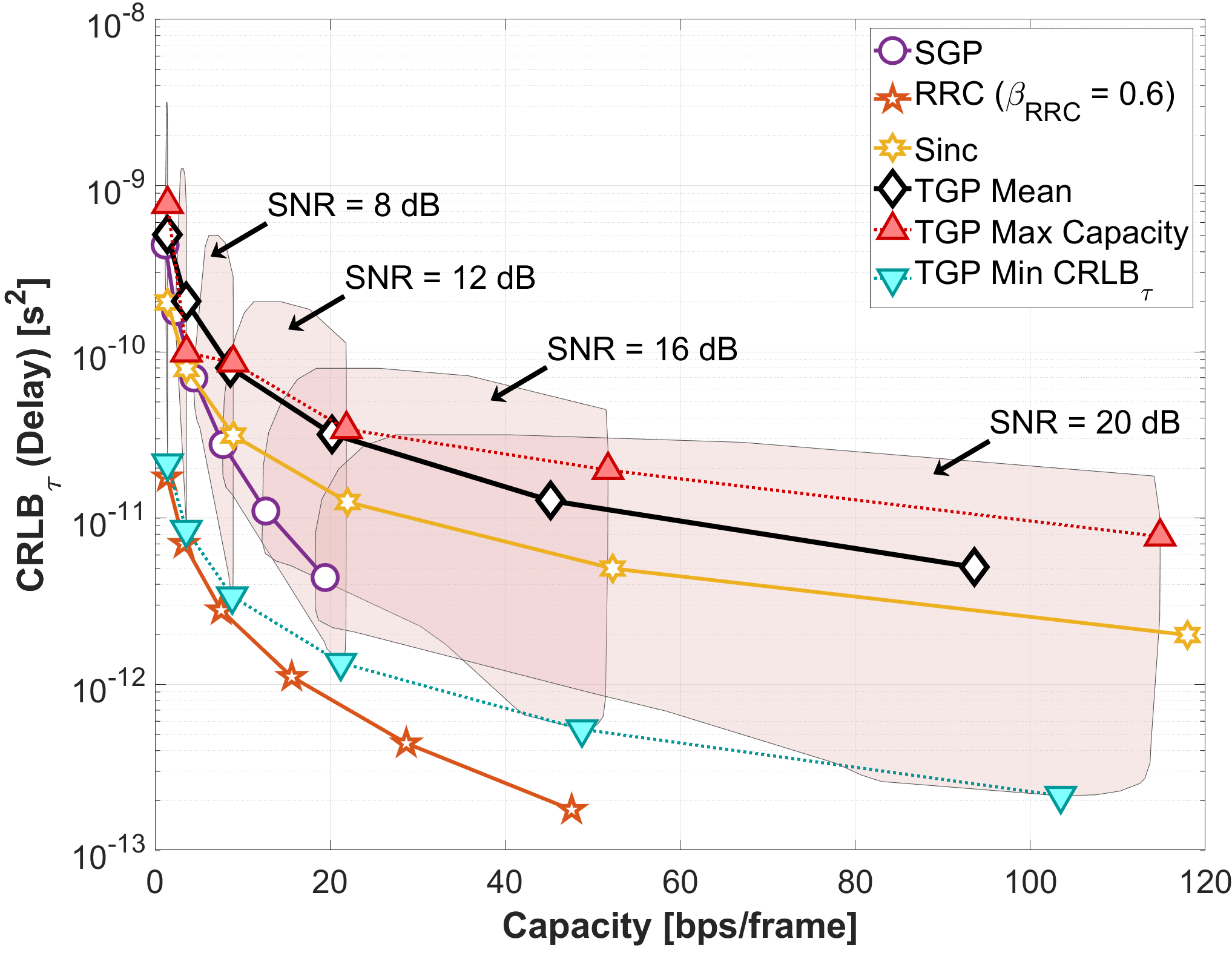}
        \caption{Trade-off: Capacity vs. Delay \gls{crlb}}
        \label{fig:tradeoff_cap_vs_crb_tau}
    \end{subfigure}
    \vspace{0.5cm}   
    \begin{subfigure}{0.48\textwidth}
        \centering
        \includegraphics[width=0.95\textwidth]{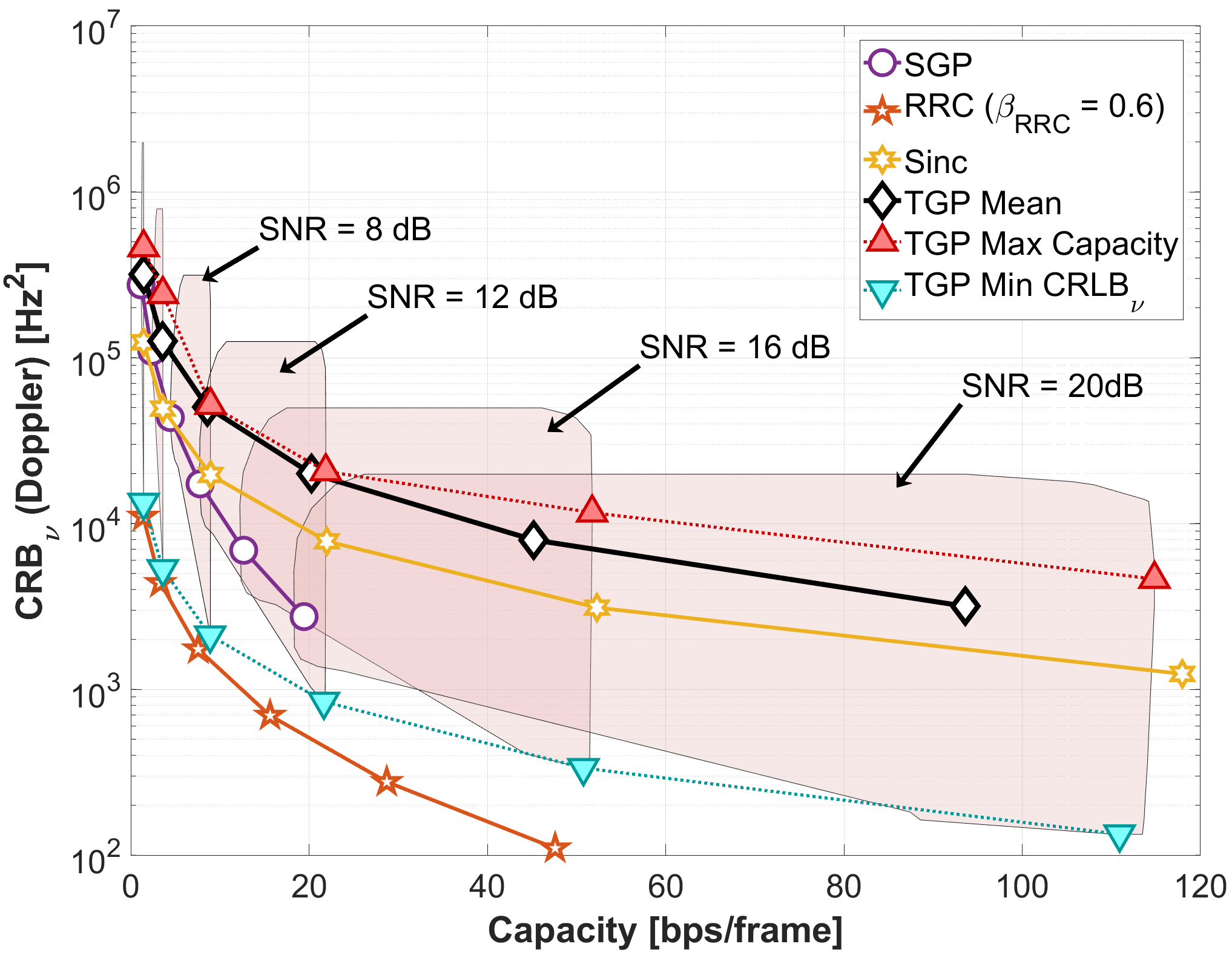}
        \caption{Trade-off: Capacity vs. Doppler \gls{crlb}}
        \label{fig:tradeoff_cap_vs_crb_nu}
    \end{subfigure}   
    \caption{\small Achievable \gls{isac} trade-off region (Capacity vs. \gls{crlb}), based on parameters in Table~\ref{tab:sim_parameters}. The solid lines trace the benchmark and \gls{tgp} mean performance across the \gls{snr} range, while the shaded region illustrates the \gls{tgp} envelope.}
    \vspace{-1em}
    \label{fig:isac_tradeoff} 
\end{figure*}

We synthesize the metrics from the preceding analyses to characterize the fundamental \gls{isac} trade-off, with results presented in Fig.~\ref{fig:isac_tradeoff}. This figure plots communication capacity (x-axis) against sensing \gls{crlb} (y-axis) for both delay (Fig.~\ref{fig:tradeoff_cap_vs_crb_tau}) and Doppler (Fig.~\ref{fig:tradeoff_cap_vs_crb_nu}). The benchmarks appear as fixed operating points, while the \gls{tgp} is characterized by its complete operational envelope (derived from the 8,000 configurations), illustrating the achievable performance region with dedicated trajectories plots for communication-optimized (Max Capacity), sensing-optimized (Min \gls{crlb}), and mean.

Figure~\ref{fig:isac_tradeoff} provides compelling evidence of the \gls{tgp}'s utility. As confirmed by prior analyses, the benchmarks occupy isolated, specialized operating points: the Sinc pulse achieves high capacity (118.0~bps/frame at 20~dB) but exhibits poor sensing performance (1.98e-12~s$^2$), while the \gls{rrc} pulse achieves superior sensing (1.76e-13~s$^2$) at the expense of reduced capacity (47.6~bps/frame). The \gls{sgp} benchmark remains non-competitive in both dimensions.

The \gls{tgp} (shaded region) uniquely bridges this performance gap. A critical observation is the asymmetric nature of the communication-sensing trade-off. The communication-optimized \gls{tgp} configuration achieves maximum capacity (114.8~bps/frame) but incurs substantial sensing degradation, performing 44x worse than \gls{rrc} in delay \gls{crlb}. Conversely, the sensing-optimized \gls{tgp} achieves a \gls{crlb} of $2.14\mathrm{e}{-13}$~s$^2$, approaching \gls{rrc} performance. Notably, this sensing enhancement requires minimal capacity sacrifice: at 103.5~bps/frame, it retains 90.1\% of the \gls{tgp}'s maximum throughput. This demonstrates that the \gls{tgp} can achieve near-optimal sensing while forfeiting only marginal communication resources, occupying a balanced operating region inaccessible to any fixed benchmark. The TGP mean curve exhibits moderate performance in capacity while revealing that achieving optimal trade-off requires careful parameter tuning.
\vspace{-0.1em}
\section{Conclusion}
\label{sec:Conclusion}
In this paper, we addressed the critical need for an analytically tunable \gls{dd}-domain pulse shape to navigate the fundamental \gls{isac} trade-off by proposing the \gls{tgp}. We demonstrated that the \gls{tgp}'s parameters grant explicit and separable control over sensing precision (quantified by the \gls{crlb}) and communication performance (measured by ergodic capacity). Our analysis revealed a powerful decoupling mechanism: the envelope parameter dictates received power, while the phase parameters exclusively reshape the \gls{isi} structure, with $\beta_c$ emerging as the dominant mechanism for \gls{isi} mitigation and channel near-diagonalization. For sensing, our closed-form \gls{fim} derivation identified the explicit conditions for achieving estimator decoupling. Finally, our comprehensive trade-off analysis demonstrated that the \gls{tgp} navigates a continuous performance space connecting the specialized extremes of benchmark pulses, approaching Sinc's communication capacity and \gls{rrc}'s sensing precision. Critically, we uncovered an asymmetric trade-off nature where near-optimal sensing performance is achieved while sacrificing less than 10\% of maximum communication throughput, revealing a favorable operational regime unattainable by any static pulse design.

While our findings establish a strong theoretical foundation, they assume perfect \gls{csir} and do not fully characterize hardware implementation costs. Future work should investigate the \gls{tgp}'s robustness to practical impairments, evaluate performance under imperfect \gls{csi}, and explore adaptive algorithms that dynamically tune the \gls{tgp} parameters in real-time to optimize the \gls{isac} trade-off for changing channel conditions and application demands.
%
\appendix
%
\vspace{-0.5em}
\subsection{Analytical Derivation of the Fisher Information Matrix}
\label{app:fim_derivation}

This appendix provides a step-by-step analytical derivation of the intrinsic \gls{fim} elements for the unit-energy TGP waveform. The derivation starts from the Slepian-Bangs formula~\cite{Slepian1954} and proceeds by showing the analytical integration of the derivative terms.

\vspace{0.1cm}
\subsubsection{Step 1: Partial Derivatives of the Mean Echo Signal}
To compute the intrinsic \gls{fim} elements, we evaluate the partial derivatives of the mean echo signal $\mu$ with respect to $\tau_T$ and $\nu_T$ at the origin (i.e., true parameters set to zero)~\cite{Dogandzic2001,Kay1993}. Applying the derivative operators to \eqref{eq:mean_echo_signal} yields the auxiliary functions $A(\tau, \nu) \triangleq \frac{\partial \mu}{\partial \tau_T}\big|_{\mathbf{0}}$ and $B(\tau, \nu) \triangleq \frac{\partial \mu}{\partial \nu_T}\big|_{\mathbf{0}}$ defined as:
\begin{align}
    A(\tau, \nu) &= -\frac{\partial x_{\text{dd}}}{\partial \tau} + j2\pi\nu \cdot x_{\text{dd}}, \label{eq:A_def} \\
    B(\tau, \nu) &= -\frac{\partial x_{\text{dd}}}{\partial \nu} - j2\pi\tau \cdot x_{\text{dd}}. \label{eq:B_def}
\end{align}
For the \gls{tgp} waveform in \eqref{eq:tgp_waveform_explicit}, computing these derivatives analytically and substituting into \eqref{eq:A_def}--\eqref{eq:B_def} leads to the explicit forms used for integration:
\begin{align}
    A(\tau, \nu) &= \pi \left( \frac{4\tau}{\gamma T^2} - j \left[\frac{2\alpha_c\tau}{T^2} + (\beta_c - 2)\nu\right] \right) x_{\mathrm{dd}}(\tau, \nu), \label{eq:A_final} \\
    B(\tau, \nu) &= \pi \left( \frac{4\gamma\nu}{B^2} - j\tau(\beta_c + 2) \right) x_{\mathrm{dd}}(\tau, \nu). \label{eq:B_final}
\end{align}

\vspace{0.1cm}
\subsubsection{Step 2: \gls{fim} Element Integration}
With the derivative terms $A(\tau, \nu)$ and $B(\tau, \nu)$ established, we now compute the elements of the intrinsic \gls{fim}, $I^{(0)}$. This requires solving the integrals of their respective power and cross-power terms. A key insight for this calculation is that the \gls{fim} integrals reduce to the statistical moments of the waveform's energy distribution in the \gls{dd}-plane. This connection is fundamental in radar theory: the \gls{fim} elements represent the curvature of the ambiguity function at the origin, which in turn is governed by the waveform's second-order energy moments~\cite{Sussman1962}. For the unit-energy \gls{tgp}, these moments (variances) are given by~\cite{Ozaktas2001}:
\begin{align}
\sigma_\tau^2 &\triangleq \iint_{-\infty}^{\infty} \tau^2 |x_{\mathrm{dd}}(\tau, \nu)|^2 \,d\tau d\nu = \frac{\gamma T^2}{4\pi}, \\
\sigma_\nu^2 &\triangleq \iint_{-\infty}^{\infty} \nu^2 |x_{\mathrm{dd}}(\tau, \nu)|^2 \,d\tau d\nu = \frac{B^2}{4\pi\gamma}.
\end{align}
The covariance moment $\iint \tau\nu |x_{\mathrm{dd}}|^2$ is zero due to the symmetry of the waveform envelope.

\paragraph{Calculation of $I^{(0)}_{\nu\nu}$}
The intrinsic \gls{fim} element for Doppler, $I^{(0)}_{\nu\nu}$, corresponds to the integral of $|B(\tau, \nu)|^2$, scaled appropriately according to the Slepian-Bangs definition for real parameters of complex signals~\cite{Kay1993}. The integrand is first expanded from \eqref{eq:B_final}:
\begin{align}
    |B(\tau, \nu)|^2 \!\!&= \!\!\left[ \frac{16\pi^2\gamma^2}{B^4}\nu^2 \!\!+ \!\pi^2(\beta_c+2)^2 \tau^2 \right] |x_{\mathrm{dd}}(\tau, \nu)|^2.
\end{align}
Integrating this expression term-by-term using the pulse shaping variances $\sigma_\nu^2$ and $\sigma_\tau^2$ and applying the factor of $1/2$ prescribed by the Slepian-Bangs formula for real-valued parameters in complex baseband signals~\cite{Kay1993} yields the final result:
\begin{align}
    I^{(0)}_{\nu\nu} \!\!&= \frac{1}{2} \left[ \frac{16\pi^2\gamma^2}{B^4}\sigma_\nu^2 + \pi^2(\beta_c+2)^2\sigma_\tau^2 \right] \nonumber \\
    &= \frac{\pi\gamma}{8B^2} \Big( 16 + B^2 T^2 (\beta_c+2)^2 \Big).
\end{align}
This result can be factorized into the compact form presented in the main text.

\paragraph{Calculation of $I^{(0)}_{\tau\tau}$}
Following an analogous procedure for the delay-domain element, we integrate the power of the derivative term, $|A(\tau, \nu)|^2$. The expansion of the integrand from \eqref{eq:A_final} yields multiple terms:
\begin{align}
    |A(\tau, \nu)|^2 &= \pi^2 \!\! \left[ \left(\frac{4\tau}{\gamma T^2}\right)^2 \!\!\!+\!\! \left(\frac{2\alpha_c\tau}{T^2} + (\beta_c-2)\nu\right)^2 \right] \!|x_{\mathrm{dd}}|^2 \nonumber \\
    &\hspace{-0.5cm} = \pi^2 \left[ C_1\tau^2 + C_2\tau^2 + C_3\tau\nu + C_4\nu^2 \right] |x_{\mathrm{dd}}|^2,
\end{align}
where
\begin{equation}
    C_1 \!=\! \frac{16}{\gamma^2 T^4}, \; C_2 \!=\! \frac{4\alpha_c^2}{T^4}, \; C_3 \!=\! \frac{4\alpha_c(\beta_c-2)}{T^2}, \; C_4 \!=\! (\beta_c-2)^2. \label{eq:C_coeffs}
\end{equation}
Upon integration, the cross-term containing $\tau\nu$ evaluates to zero due to odd symmetry. The remaining terms are integrated using the known variances and scaled by the requisite factor of $1/2$ \cite{Kay1993}, we have:
\begin{align}
    I^{(0)}_{\tau\tau} &= \frac{1}{2} \left[ \pi^2 \left(\frac{16}{\gamma^2 T^4} + \frac{4\alpha_c^2}{T^4}\right)\sigma_\tau^2 + \pi^2(\beta_c-2)^2\sigma_\nu^2 \right] \nonumber \\
    &= \frac{\pi}{8\gamma T^2} \Big( 4(4+\alpha_c^2\gamma^2) + B^2 T^2 (\beta_c-2)^2 \Big).
\end{align}
\paragraph{Calculation of $I^{(0)}_{\tau\nu}$}
Finally, the coupling term is found by integrating the real part of the cross-product, $A(\tau,\nu)^* B(\tau,\nu)$. The full expansion of the product's real part contains several terms:
\begin{align}
    &\text{Re}\{A^*B\} = \pi^2 \Bigg[ \frac{16\tau\nu}{\gamma T^2 B^2} + \frac{2\alpha_c(\beta_c+2)}{T^2}\tau^2 \nonumber \\
    & \hspace{3.0cm} + \frac{(\beta_c-2)(\beta_c+2)}{1}\tau\nu \Bigg] |x_{\mathrm{dd}}|^2.
\end{align}
Upon integration, all terms containing an odd power of $\tau$ or $\nu$ (i.e., the $\tau\nu$ terms) evaluate to zero due to symmetry. Only the term proportional to $\tau^2$ remains. Applying the necessary scaling factor of $1/2$ \cite{Kay1993}, the integral becomes:
\begin{align}
    I^{(0)}_{\tau\nu} &= \frac{1}{2} \iint_{-\infty}^{\infty} \pi^2\left(\frac{2\alpha_c(\beta_c+2)}{T^2}\tau^2\right)|x_{\mathrm{dd}}|^2\,d\tau d\nu \nonumber \\
    &= \frac{\pi}{4}\alpha_c\gamma(2+\beta_c). \label{eq:Icross_final_factored}
\end{align}
This completes the derivation of the intrinsic \gls{fim} elements for the unit-energy \gls{tgp} waveform.

\subsection{Covariance Matrix Coefficients}
\label{app:covariance_coeffs}

We explicitly detail the complex coefficients $\{A_\tau, B_\tau, A_\nu, B_\nu, C_0\}$ defining the separable quadratic phase in \eqref{eq:quadratic_phase_general}. Defining the discrete grid displacements as $d_k = k-m$, $d'_k = k'-m$, $d_l = l-n$, and $d'_l = l'-n$, the purely real quadratic coefficients are:
\begin{align}
    A_\tau = -\frac{4\pi}{\gamma T^2}, \quad \quad A_\nu = -\frac{4\pi\gamma}{B^2}. 
\end{align}
The complex linear coefficients capture the interaction between displacements and pulse parameters:
\begin{align}
    B_\tau &= \frac{4\pi \Delta\tau}{\gamma T^2}(d_k + d'_k) \nonumber \\
    &\quad - j \pi \left[ \frac{2\alpha_c\Delta\tau}{T^2}(d_k - d'_k) + \beta_c\Delta\nu(d_l - d'_l) \right], \label{eq:B_tau_final} \\
    B_\nu &= \frac{4\pi\gamma\Delta\nu}{B^2}(d_l + d'_l) \nonumber \\
    &\quad + j \pi \left[ \frac{2}{N}(d_l - d'_l) - \beta_c\Delta\tau(d_k - d'_k) \right]. \label{eq:B_nu_final}
\end{align}
Finally, the constant term $C_0$ represents the total phase/amplitude offset:
\begin{align}
    C_0 &=  - 2\pi \left[ \frac{\Delta\tau^2}{\gamma T^2}(d_k^2 + {d'_k}^2) + \frac{\gamma \Delta\nu^2}{B^2}(d_l^2 + {d'_l}^2) \right] \nonumber \\
        &\hspace{-0.4cm}+ j \pi \left[ \frac{\alpha_c \Delta\tau^2}{T^2}(d_k^2 - {d'_k}^2) + \beta_c \Delta\tau \Delta\nu (d_k d_l - d'_k d'_l) \right]. \label{eq:C_0_final}
\end{align}

\vspace{-.1mm}
\bibliographystyle{IEEEtran}
\bibliography{refs} 

\end{document}